\documentclass[aps,prd,10pt,onecolumn,nofootinbib]{revtex4-2}

\usepackage[T1]{fontenc}
\usepackage[utf8]{inputenc}
\usepackage{amsmath,amssymb,bm,mathtools,mathrsfs}
\usepackage{graphicx}
\usepackage{booktabs}
\usepackage[section]{placeins}
\usepackage{hyperref}
\usepackage{tikz}
\usetikzlibrary{arrows.meta,positioning,calc}
\hypersetup{hidelinks,pdftitle={Light-front diagnostics in the 't Hooft model: II. Boundary cancellations, ERBL spectral sums, and analyticity of the EMT form factor},pdfauthor={Arkadiy I. Syamtomov}}

\newcommand{\mtilde}{\widetilde m}
\newcommand{\dd}{\mathrm d}
\newcommand{\MD}{\mathcal M_{2,n}^{\mathrm D}}
\newcommand{\ME}{\mathcal M_{2,n}^{\mathrm E}}
\newcommand{\Mfull}{\mathcal M_{2,n}}
\newcommand{\Jodd}{J_n^-}
\newcommand{\Gproj}{\mathcal G_n}
\newcommand{\FP}{\operatorname{FP}}
\newcommand{\PV}{\operatorname{PV}}
\newcommand{\res}{\operatorname*{Res}}
\begin{document}

\title{Light-front diagnostics in the 't~Hooft model: II. Boundary cancellations, ERBL spectral sums, and analyticity of the EMT form factor}

\author{Arkadiy I. Syamtomov}
\affiliation{Bogolyubov Institute for Theoretical Physics\\National Academy of Sciences of Ukraine\\14-B Metrolohichna Street, 03143 Kyiv, Ukraine}

\begin{abstract}\noindent
A diagonal light-front overlap contains near-forward energy--momentum-tensor (EMT) information but is not itself a local matrix element.  In the large-$N_c$ 't~Hooft model we complete the second GPD moment with its ERBL contribution and follow how the two support regions combine.  Boundary exponents generate fractional powers and, at $\beta=1/2$, resonant logarithms in the separate terms.  We show that these structures cancel, including the interference of the leading and next indicial families.  At the resonance, cancellation of the double and single logarithms also yields an inverse-mass-squared residue sum rule; above it, the two regions combine into a finite fourth-order coefficient.  The complete EMT form factor is therefore analytic through curvature order.  The ERBL sector leaves the slope unchanged but is essential for curvature and higher derivatives.  We extract stable canonical and heavy curvatures, estimates with systematic envelopes for the second and third light excitations, and the longitudinal trace radius.  The intermediate-state decomposition further shows how constituent mass and excitation redistribute pole strength and generate removable poles and interference zeros.
\end{abstract}

\maketitle

\section{Introduction}
\label{sec:intro}

Light-front descriptions of confined systems raise the question of how much of a hadron's energy--momentum structure is already encoded in its leading Fock-sector wave function.  Forward observables often admit direct wave-function representations, but nonzero momentum transfer probes a more restrictive issue.  A diagonal overlap preserves particle number, whereas a local current or energy--momentum-tensor (EMT) matrix element can also receive contributions from pair creation and intermediate bound states.  The large-$N_c$ 't~Hooft model provides a particularly controlled setting in which this distinction can be followed from the bound-state equation to the complete off-forward amplitude.

Part~I compared four benchmark mass families and isolated the information carried by the leading quark--antiquark wave function~\cite{Syamtomov:PartI}.  Here we use its light--light, equal-mass reference, and heavy--heavy sectors, denoted below as the light, canonical, and heavy benchmarks.  For the diagonal GPD overlap, the term linear in the asymmetric skewness variable $b$ vanishes, while the regular $b^2$ and $b^3$ coefficients coincide.  Their state dependence is governed by the positive weighted gradient norm
\begin{equation}
 I_n=\int_0^1\dd x\,x(1-x)[\phi_n'(x)]^2.
 \label{eq:introgradient}
\end{equation}
The same analysis made the limitation of the diagonal overlap explicit.  At the reference mass $\mtilde^{\,2}=1$, where $\beta=1/2$, its boundary expansion contains $b^4\ln^2(1/b)$.  Such a support-dependent term cannot occur in the Taylor expansion of the complete massive EMT form factor about $t=0$.

Part~I separated the forward Hamiltonian expectation value into the explicit constituent-mass term and the instantaneous confining interaction.  This is not the four-dimensional separation of renormalized quark and gluon EMT traces, whose individual operators mix and whose partition is scheme dependent~\cite{Hatta:2018ina}.  In strict $1+1$ dimensions $T_g^{++}=0$, so the second quark GPD moment equals the total $T^{++}$ matrix element~\cite{Ji:2021}.  The DGLAP--ERBL division below is therefore a support decomposition of the same quark bilocal operator, not quark--gluon operator mixing.

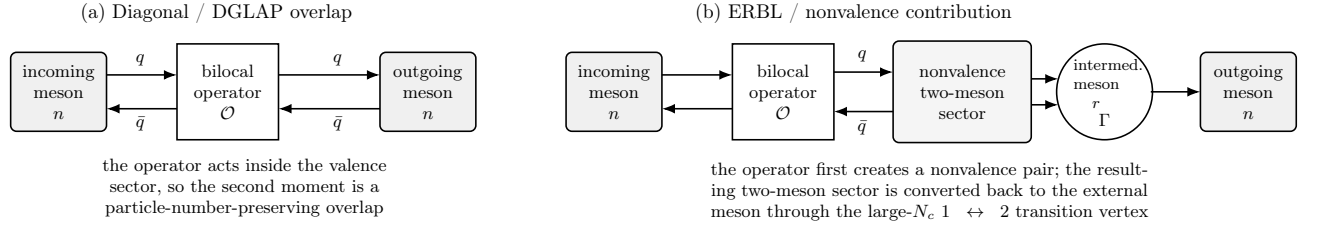
\begin{figure}[t]
\centering
\resizebox{0.95\textwidth}{!}{%
\begin{tikzpicture}[x=1cm,y=1cm,>=Latex,thick,font=\small]
  \node[font=\normalsize] at (3.8,3.45) {(a) Diagonal / DGLAP overlap};
  \draw[rounded corners=3pt,fill=gray!12] (0.15,1.35) rectangle (1.85,2.75);
  \node[align=center,text width=1.35cm] at (1.0,2.05) {incoming\\ meson\\ $n$};
  \draw[rounded corners=3pt,fill=gray!12] (6.7,1.35) rectangle (8.4,2.75);
  \node[align=center,text width=1.35cm] at (7.55,2.05) {outgoing\\ meson\\ $n$};
  \draw[fill=white] (3.1,1.2) rectangle (4.9,2.9);
  \node[align=center,text width=1.4cm] at (4.0,2.05) {bilocal\\ operator\\ $\mathcal O$};
  \draw[->] (1.85,2.35) -- (3.1,2.35);
  \draw[<-] (1.85,1.75) -- (3.1,1.75);
  \draw[->] (4.9,2.35) -- (6.7,2.35);
  \draw[<-] (4.9,1.75) -- (6.7,1.75);
  \node at (2.45,2.62) {$q$};
  \node at (2.45,1.48) {$\bar q$};
  \node at (5.95,2.62) {$q$};
  \node at (5.95,1.48) {$\bar q$};
  \node[align=center,text width=7.4cm] at (4.3,0.35) {the operator acts inside the valence sector, so the second moment is a particle-number-preserving overlap};

  \node[font=\normalsize] at (15.1,3.45) {(b) ERBL / nonvalence contribution};
  \draw[rounded corners=3pt,fill=gray!12] (10.0,1.35) rectangle (11.7,2.75);
  \node[align=center,text width=1.35cm] at (10.85,2.05) {incoming\\ meson\\ $n$};
  \draw[fill=white] (12.95,1.2) rectangle (14.75,2.9);
  \node[align=center,text width=1.4cm] at (13.85,2.05) {bilocal\\ operator\\ $\mathcal O$};
  \draw[rounded corners=3pt,fill=gray!8] (15.8,1.15) rectangle (18.25,2.95);
  \node[align=center,text width=2.0cm] at (17.03,2.05) {nonvalence\\ two-meson\\ sector};
  \draw[fill=white] (19.55,2.05) circle (0.85);
\node[
    align=center,
    font=\footnotesize,
    text width=0.80cm
] at (19.40,2.20)
{intermed.\\meson $r$};
\node[
    font=\small
] at (19.55,1.55)
{$\Gamma$};  
  \draw[rounded corners=3pt,fill=gray!12] (21.25,1.35) rectangle (22.95,2.75);
  \node[align=center,text width=1.35cm] at (22.10,2.05) {outgoing\\ meson\\ $n$};
  \draw[->] (11.7,2.35) -- (12.95,2.35);
  \draw[<-] (11.7,1.75) -- (12.95,1.75);
  \draw[->] (14.75,2.40) -- (15.8,2.40);
  \draw[<-] (14.75,1.70) -- (15.8,1.70);
  \node at (15.25,2.68) {$q$};
  \node at (15.25,1.42) {$\bar q$};
  \draw[->] (18.25,2.28) -- (18.73,2.28);
  \draw[->] (18.25,1.82) -- (18.73,1.82);
  \draw[->] (20.37,2.05) -- (21.25,2.05);
  \node[align=center,text width=10.5cm] at (16.45,0.28) {the operator first creates a nonvalence pair; the resulting two-meson sector is converted back to the external meson through the large-$N_c$ $1\leftrightarrow2$ transition vertex};
\end{tikzpicture}%
}
\caption{Schematic light-front mechanisms contributing to the second GPD moment.  Panel (a) shows the DGLAP contribution, in which the bilocal operator acts within the valence $q\bar q$ sector and produces the diagonal overlap.  Panel (b) shows the ERBL mechanism, in which the bilocal operator creates a nonvalence pair, the intermediate state propagates through a two-meson sector, and the large-$N_c$ transition vertex $\Gamma$ reconnects that sector to the external meson through the confined intermediate meson $r$.}
\label{fig:supportschematic}
\end{figure}

The local quantity that makes this tension precise is the complete second GPD moment.  Once the diagonal and pair-creation regions are combined, it is the matrix element of the plus--plus component of the EMT~\cite{Ji:1997DVCS,Diehl:2003}; in the 't~Hooft model it reduces to the unique spin-zero EMT form factor in $1+1$ dimensions~\cite{Ji:2021}.  Its analyticity below the first physical $t$-channel singularity requires the fractional powers and logarithms found in the separate light-front contributions to cancel in their sum.  The problem is therefore not to reinterpret the diagonal overlap, but to determine the missing nonvalence term and to resolve the higher-order boundary structure of the diagonal contribution with the same subtraction prescription.  Figure~\ref{fig:supportschematic} summarizes the physical content of these two support regions.

Complete large-$N_c$ form factors and their confined-meson pole representation are known~\cite{Einhorn:1976,Callan:1976,Burkardt:2000}, as is the DGLAP--ERBL GPD~\cite{Burkardt:2000,Jia:2024} and the two-dimensional EMT decomposition~\cite{Ji:2021}.  To our knowledge, however, the near-forward local second moment has not previously been resolved at the level of its separate support regions.  We project the pole tower with the second-moment source and combine it with the nonuniform boundary expansion of the diagonal overlap.

The required boundary theory is not new.  Hildebrandt established the global variational and spectral properties and the singular-boundary behavior, Lewy proved a convergent fractional-power expansion near the singularity, and Abe organized the finite-interval wave function into families associated with all positive roots of the indicial equation~\cite{Federbush:1977,Hildebrandt:1978,Hildebrandt:1979,Lewy:1979,Abe:1999}.  We use this multi-root expansion to determine how the first descendants enter the near-forward diagonal overlap.  In that overlap, matched boundary--interior subtraction removes the apparent $b^{2+2\beta}$ term.  In the ERBL contribution, reflection symmetry selects the odd intermediate spectrum.  The first surviving correction is then different in the three regimes $0<\beta<1/2$, $\beta=1/2$, and $1/2<\beta<1$.
The near-forward power counting is summarized by
\begin{equation*}
\begin{array}{c|ccc}
 &0<\beta<\tfrac12&\beta=\tfrac12&\tfrac12<\beta<1\\ \hline
J_n^-&b^{1+2\beta}&b^2\ln^2(1/b)&b^2\\
\mathcal M_{2,n}^{\mathrm E}&b^{3+2\beta}&b^4\ln^2(1/b)&b^4
\end{array}
\end{equation*}
before the cancellations required by the complete EMT form factor are imposed.  On the light side the leading family gives the first line of this hierarchy.  A second term from interference with the next indicial family also occurs before $b^4$; Sec.~\ref{sec:diagonalcompletion} isolates its diagonal coefficient, and Sec.~\ref{subsec:pairwise-theorem} proves its exact ERBL cancellation.

The spectral organization of the ERBL contribution supplies the complementary finite-$t$ information.  Parity decomposition and spectral-determinant methods for the 't~Hooft Hamiltonian are well developed~\cite{Fateev:2009,Ambrosino:2025,Litvinov:2025}, as is the general resolvent--minor relation~\cite{Feinberg:2004}.  We use them to separate the universal odd spectrum from source-dependent transition amplitudes, thereby identifying visible and removable poles and interference zeros.

Combining these ingredients gives a common physical picture across the three mass regimes.  In the light system, the leading and mixed boundary powers are canceled by ERBL sources that are Hamiltonian images; at $\beta=1/2$ the same mechanism cancels the resonant logarithms and produces an inverse-mass-squared sum rule; beyond the resonance the two support regions combine into a finite curvature.  The completed form factor fixes the slope and longitudinal trace radius, while its source-projected resolvent describes the finite-transfer pole content.

The strict chiral ground state is kept separate because its constant wave function makes the ERBL source vanish pointwise, as is already visible in the complete GPD kernel of Ref.~\cite{Jia:2024}, whereas the finite-mass analysis relies on a nonzero gap between $t=0$ and the first physical singularity.  Throughout, the model is used as a longitudinal field theory.  It has no transverse momentum transfer and therefore does not define impact-parameter densities, pressure, shear, or a transverse gravitational radius~\cite{Diehl:2003,Polyakov:2018}.

\section{Boundary structure of the diagonal overlap}
\label{sec:diagonalcompletion}

The obstruction identified in the Introduction originates in a narrow momentum-fraction region.  When the longitudinal transfer is small, the shifted argument of the final-state wave function approaches the boundary on the scale $x=O(b)$.  An ordinary Taylor expansion of the diagonal overlap is therefore not uniform over the full integration interval.  This requires separating the terms fixed by the regular interior of the wave function from those controlled by its small-$x$ behavior.

Following the standard dimensionless normalization of the 't~Hooft model and the convention used in Part~I~\cite{tHooft:1974,Syamtomov:PartI}, define the mass-squared scale
\begin{equation}
 \Lambda\equiv\frac{g^2N_c}{\pi},
 \qquad
 \mtilde_i^2=\frac{m_i^2}{\Lambda},
 \qquad
 M_n^2=\frac{M_{n,\mathrm{phys}}^2}{\Lambda},
 \qquad
 t=\frac{t_{\mathrm{phys}}}{\Lambda}.
 \label{eq:units}
\end{equation}
The scale $\Lambda$ is the quantity denoted by $\lambda$ in Part~I.  All Hamiltonians, pole masses, and momentum transfers below are dimensionless in units of $\Lambda$ unless physical units are restored explicitly.
For equal constituent masses, the normalized meson wave functions obey the finite-interval bound-state equation~\cite{tHooft:1974,Hildebrandt:1979,Fateev:2009}
\begin{equation}
 M_n^2\phi_n(x)=
 \frac{\mtilde^2}{x(1-x)}\phi_n(x)
 +\mathrm{PV}\!\int_0^1\dd y\,
 \frac{\phi_n(x)-\phi_n(y)}{(x-y)^2},
 \qquad
 \int_0^1\dd x\,\phi_m(x)\phi_n(x)=\delta_{mn}.
 \label{eq:thooft}
\end{equation}
The relevance of the boundary exponent follows directly from the shifted overlap: the region $x\sim b$ samples the wave function at arguments of order $b$.  With $\phi_n(x)\sim C_nx^\beta$, the mass term and singular integral both scale as $x^{\beta-1}$.  After $y=xu$, the singular coefficient contains the finite-Hilbert-transform identity
$\mathrm{PV}\!\int_0^\infty \dd u\,(1-u^\beta)/(1-u)^2=\pi\beta\cot(\pi\beta)-1$.
Matching it to the mass term gives~\cite{tHooft:1974,Abe:1999,Fateev:2009}
\begin{equation}
 \mtilde^2-1+\pi\beta\cot(\pi\beta)=0,
 \qquad
 \phi_n(x)\sim C_nx^\beta
 \quad (x\to0).
 \label{eq:beta}
\end{equation}
Thus $\beta$ is introduced here not as an additional convention, but because it controls the nonuniform terms that must eventually be matched to the ERBL sector.

Write the initial and final meson momenta with longitudinal rapidities $-\eta$ and $+\eta$.  Then $\xi=\tanh\eta$ and the asymmetric overlap variable $b=1-e^{-2\eta}=2\xi/(1+\xi)$.  Since $t=-4M_n^2\sinh^2\eta$, the exact elastic relation is~\cite{Syamtomov:PartI}
\begin{equation}
 b=\frac{2\xi}{1+\xi},
 \qquad
 \xi=\frac{b}{2-b},
 \qquad
 t=-\frac{M_n^2b^2}{1-b}.
 \label{eq:kinematics}
\end{equation}
Thus $b$ is the natural longitudinal boost variable, not a regulator: $b=\sqrt{-t}/M_n+O(t)$.  An isolated noninteger power $b^\alpha$ begins as $(-t)^{\alpha/2}$ and is nonanalytic at $t=0$.  The particle-number-preserving contribution to the second GPD moment is the diagonal overlap~\cite{Burkardt:2000,Jia:2024,Syamtomov:PartI}
\begin{equation}
 \MD(b)=
 \frac{4(1-b)}{(2-b)^2}
 \int_0^1\dd x\,
 \bigl[b+2(1-b)x\bigr]
 \phi_n(x)\phi_n\!\left(x+b(1-x)\right).
 \label{eq:MDexact}
\end{equation}
Equation~\eqref{eq:MDexact} is the particle-number-preserving part of the second moment.  At nonzero transfer it is not the complete EMT form factor.

Part~I established the forward normalization, the vanishing linear term, and the regular quadratic and cubic coefficients of this overlap~\cite{Syamtomov:PartI}.  In particular,
\begin{equation}
 \MD(0)=1,
 \qquad
 \left.\frac{\dd\MD}{\dd b}\right|_{b=0}=0.
 \label{eq:MDforward}
\end{equation}
The first equality is the longitudinal momentum sum rule for the equal-mass two-body state.  The second excludes an $O(b)\sim O(\sqrt{-t})$ term in the diagonal overlap.

Separating the regular terms fixed in Part~I, we write
\begin{equation}
 \MD(b)=1+A_{2,n}^{\mathrm D}(b^2+b^3)+R_{\beta,n}^{\mathrm D}(b),
 \qquad
 R_{\beta,n}^{\mathrm D}(b)=o(b^3),
 \label{eq:MDcontrolled}
\end{equation}
where
\begin{equation}
 A_{2,n}^{\mathrm D}
 =-\frac14-\frac12I_n,
 \qquad
 I_n=\int_0^1\dd x\,x(1-x)[\phi_n'(x)]^2.
 \label{eq:A2D}
\end{equation}
After the finite-part cancellation and the multi-root boundary hierarchy developed below are included, $R_{\beta,n}^{\mathrm D}=o(b^3)$.  The common coefficient of $b^2$ and $b^3$ is not accidental: from Eq.~\eqref{eq:kinematics}, $t=-M_n^2(b^2+b^3+\cdots)$, so a form factor analytic and linear in $t$ produces precisely this pair of terms.  The coefficient $A_{2,n}^{\mathrm D}$ becomes the physical EMT slope only after the ERBL contribution is shown to have no term of order $b^2$.

The remainder is defined by
\begin{equation}
 R_{\beta,n}^{\mathrm D}(b)=
 \MD(b)-1-A_{2,n}^{\mathrm D}(b^2+b^3)
 \label{eq:Rdiagdef}
\end{equation}
and contains the boundary-sensitive information that remains beyond the diagonal finite-part cancellation and the canonical double logarithm already obtained in Part~I.  Its mass-dependent asymptotics determine the fractional, logarithmic, or finite fourth-order term that must be combined with the ERBL contribution.

The finite-interval analyses of Refs.~\cite{Hildebrandt:1979,Lewy:1979,Abe:1999} show that an independent integer-power branch is inadmissible and that the local solution is organized by all roots of the indicial function
\begin{equation}
 F(a)=\mtilde^2-1+\pi a\cot(\pi a),
 \qquad
 F(\gamma_j)=0,
 \qquad
 j<\gamma_j<j+1,
 \qquad
 \gamma_0=\beta.
 \label{eq:gammaroots}
\end{equation}
In the present variable their result may be written
\begin{equation}
 \phi_n(x)\sim
 \sum_{j\geq0}C_{n,j}x^{\gamma_j}
 \sum_{N\geq0}c_{\gamma_j,N}^{(n)}x^N.
 \label{eq:remainderfamilies}
\end{equation}
Only the first descendant of the leading family is required for the analytic overlap coefficient derived below.  We also record the second descendant because it is needed for the recurrence-consistent composite check.  Converting the published recurrence to the normalization of Eq.~\eqref{eq:thooft} gives~\cite{Abe:1999}
\begin{equation}
 \rho_n\equiv c_{\beta,1}^{(n)}
 =\beta+\frac{M_n^2}{\pi\cot(\pi\beta)}.
 \label{eq:rho}
\end{equation}
The next leading-family descendant is fixed by the same recurrence.  At order $x^{\beta+1}$, the fractional part of the eigenvalue equation gives
\begin{equation}
 2\pi\cot(\pi\beta)c_{\beta,2}^{(n)}
 =M_n^2\rho_n-(\widetilde m^2-1)(1+\rho_n).
 \label{eq:cbeta2recurrence}
\end{equation}
Using Eq.~\eqref{eq:rho} and the indicial condition then yields
\begin{equation}
 c_{\beta,2}^{(n)}
 =\frac{M_n^2\rho_n-(\widetilde m^2-1)(1+\rho_n)}{2\pi\cot(\pi\beta)}
 =\frac{\beta+\rho_n^2}{2}.
 \label{eq:cbeta2}
\end{equation}
Thus $c_{\beta,2}^{(n)}$ is not a new matching parameter.  The amplitudes of the remaining root families are global matching data and are not additional local integration constants~\cite{Hildebrandt:1979,Lewy:1979,Abe:1999}.

Part~I derived the matched finite-part cancellation generated by the product of the two leading boundary powers in the light regime $0<\beta<1/2$~\cite{Syamtomov:PartI}.  The same total-derivative argument extends, with the identical Hadamard subtraction prescription, to the nonresonant post-canonical range $1/2<\beta<1$; Appendix~\ref{app:finiteparts} records this extension.  We therefore use
\begin{equation}
 \FP\!\int_0^\infty\dd u\,
 (1+2u)u^\beta(1+u)^\beta=0,
 \qquad 0<\beta<1,
 \qquad \beta\ne\frac12.
 \label{eq:Jbetazero}
\end{equation}
For $0<\beta<1/2$, this removes the candidate fractional term $b^{2+2\beta}$ between the regular $b^2$ and $b^3$ contributions; for $1/2<\beta<1$, the same identity removes its post-resonant continuation before the fourth-order coefficient is extracted.

The first nonvanishing boundary correction in the light regime comes from the interference of $x^\beta$ with its first descendant $\rho_nx^{\beta+1}$.  To obtain it directly, set $x=bu$ in the boundary layer and insert $\phi_n(x)=C_nx^\beta(1+\rho_nx)+\cdots$ into Eq.~\eqref{eq:MDexact}.  Through relative order $b$ one finds
\begin{align}
 \MD\big|_{x=O(b)}={}&C_n^2b^{2+2\beta}\FP\!\int_0^\infty\!\dd u\,
 u^\beta(1+u)^\beta\Bigl[(1+2u)+b\,Q_{\beta,n}(u)+O(b^2)\Bigr],\\
 Q_{\beta,n}(u)={}&\rho_n(1+2u)^2
 -\frac{\beta u(1+2u)}{1+u}-2u.
 \label{eq:diagboundaryexpanded}
\end{align}
The extension of the scaled integral to $u=\infty$ denotes the matched boundary finite part obtained after subtracting the regular interior Taylor expansion; it does not extend the local wave-function approximation outside its domain.  The leading finite part is Eq.~\eqref{eq:Jbetazero}.  The purely kinematic order-$b$ integrand is a total derivative,
\begin{equation}
 u^\beta(1+u)^\beta
 \left[\frac{\beta u(1+2u)}{1+u}+2u\right]
 =\frac{\dd}{\dd u}\left[u^{\beta+2}(1+u)^\beta\right],
 \label{eq:diagkinematictotalderivative}
\end{equation}
and its analytically continued finite part therefore vanishes,
\begin{equation}
 \FP\!\int_0^\infty\!\dd u\,u^\beta(1+u)^\beta
 \left[\frac{\beta u(1+2u)}{1+u}+2u\right]=0,
 \label{eq:diagkinematiczero}
\end{equation}
so only the descendant interference remains.  Analytic continuation of the three beta integrals gives
\begin{align}
 \mathcal I_\beta&\equiv\FP\!\int_0^\infty\!\dd u\,
 u^\beta(1+u)^\beta(1+2u)^2\\
 &=B(\beta+1,-2\beta-1)+4B(\beta+2,-2\beta-2)
 +4B(\beta+3,-2\beta-3)\\
 &=-\frac{\Gamma(1+\beta)\Gamma(1-2\beta)}
 {2(2\beta+1)(2\beta+3)\Gamma(1-\beta)}.
 \label{eq:Ibetadiagonal}
\end{align}
The resulting diagonal coefficient is
\begin{equation}
 \boxed{B_{\beta,n}^{\mathrm D}=C_n^2\rho_n\mathcal I_\beta
 =-\frac{C_n^2\rho_n\,\Gamma(1+\beta)\Gamma(1-2\beta)}
 {2(2\beta+1)(2\beta+3)\Gamma(1-\beta)}.}
 \label{eq:BDlight}
\end{equation}
For the representative light-mass case, independent cutoff and tail-extrapolation calculations reproduce the value obtained from the analytic beta-function expression, $\mathcal I_\beta=-0.1330052208$, within $4.2\times10^{-7}$.  This provides a numerical closure check of the derivation.

Combining this coefficient with the regular terms gives
\begin{equation}
 \MD(b)=1+A_{2,n}^{\mathrm D}(b^2+b^3)
 +B_{\beta,n}^{\mathrm D}b^{3+2\beta}
 +o\!\left(b^{3+2\beta}\right).
 \label{eq:MDlightcomplete}
\end{equation}
Equation~\eqref{eq:MDlightcomplete} contains the first nonanalytic term, but the multi-root expansion also contains mixed products of distinct boundary families.  To isolate them, retain two powers $C_px^p$ and $C_qx^q$ in the exact diagonal overlap.  The boundary rescaling $x=bu$ gives
\begin{equation}
 \left.\MD(b)\right|_{\{p,q\}}
 =C_pC_q\,\mathcal K(p,q)\,b^{2+p+q}
 +o\!\left(b^{2+p+q}\right),
 \label{eq:pairdiag}
\end{equation}
where the matched finite-part coefficient is
\begin{align}
 \mathcal K(p,q)
 &\equiv\FP\!\int_0^\infty\!\dd u\,(1+2u)
 \left[u^p(1+u)^q+u^q(1+u)^p\right]\notag\\
 &=\mathscr H(p,q)+\mathscr H(q,p),
 \qquad
 \mathscr H(p,q)=\frac{q-p}{p+q+2}B(p+1,-p-q-1).
 \label{eq:pairkernel}
\end{align}
The second equality follows term by term from beta-function recurrence:
$B(p+1,-p-q-1)+2B(p+2,-p-q-2)=\mathscr H(p,q)$.
It shows that the mixed coefficient is not merely allowed by power counting; it is explicitly generated whenever both global amplitudes are nonzero and $\mathcal K(p,q)\ne0$.

The corresponding ERBL coefficient and the pairwise closure theorem are derived in Sec.~\ref{subsec:pairwise-theorem}, after the complete source and odd-sector resolvent have been introduced; the operator identity is proved in Appendix~\ref{app:pairwise}.

For the first independent family, set $p=\beta$ and $q=\gamma_1$, with $F(\gamma_1)=0$ and $1<\gamma_1<2$.  The diagonal term is
\begin{equation}
 \MD(b)\supset B_{\beta\gamma_1,n}^{\mathrm D}b^{2+\beta+\gamma_1},
 \qquad
 B_{\beta\gamma_1,n}^{\mathrm D}
 =C_nC_{n,1}\mathcal K(\beta,\gamma_1).
 \label{eq:nextrootlight}
\end{equation}
For the representative light mass $\mtilde^2=0.04$,
\begin{equation}
 \beta\simeq0.10983,
 \qquad \gamma_1\simeq1.43313,
 \qquad \mathcal K(\beta,\gamma_1)\simeq-0.171225.
 \label{eq:lightrootsnumeric}
\end{equation}
Thus the kernel coefficient is nonzero.  The mixed term is present for any state with $C_{n,1}\neq0$; the composite fits in Appendix~\ref{app:numerics} are consistent with a nonzero amplitude, but they are not used as a precision determination of $C_{n,1}$.  Its power satisfies
\begin{equation}
 3+2\beta\simeq3.21965
 <2+\beta+\gamma_1\simeq3.54296
 <4,
 \label{eq:lightpowerordering}
\end{equation}
and Sec.~\ref{subsec:pairwise-theorem} supplies an equal ERBL term with the opposite sign.  Under the convergent multi-root expansion established in Refs.~\cite{Hildebrandt:1978,Hildebrandt:1979,Lewy:1979,Abe:1999}, descendants occur in integer steps and the positive roots are ordered as $\beta<\gamma_1<\gamma_2<\cdots$.  The next leading-family descendant has exponent $2+\beta+(\beta+2)>4$, the next independent root obeys $\gamma_2>2$, and a pair of $\gamma_1$ branches begins at $2+2\gamma_1>4$.  These assumptions exhaust the asymptotic pairings below fourth order.  Hence no fractional contribution remains below $b^4$; the finite light-sector coefficient is subsequently estimated from the direct complete-EMT-form-factor calculation in Sec.~\ref{sec:matching}.
At $\beta=1/2$, the first descendant $x^{\beta+1}=x^{3/2}$ has the same power as the next indicial solution.  We use the resonant boundary and diagonal-overlap results established in Part~I~\cite{Syamtomov:PartI},
\begin{align}
 \phi_n(x)&=C_nx^{1/2}+D_nx^{3/2}\ln x+K_nx^{3/2}+\cdots,
 &D_n&=-\frac{2M_n^2}{3\pi^2}C_n,
 \label{eq:canonicalendpoint}\\
 \MD(b)&\supset d_{2,n}b^4\ln^2\!\frac1b,
 &d_{2,n}&=\frac{C_nD_n}{64}.
 \label{eq:canonicaldiagonaldoublelog}
\end{align}
For $1/2<\beta<1$, the same finite-part cancellation removes the $b^{2+2\beta}$ term, while the descendant contribution $b^{3+2\beta}$ starts beyond fourth order.  The expansion therefore reaches $b^4$ without an intervening fractional power; Sec.~\ref{sec:matching} derives its finite coefficient.

The fractional-power hierarchy itself is established in the earlier boundary literature.  Its use in the near-forward overlap gives the closed coefficient $B_{\beta,n}^{\mathrm D}$ for $0<\beta<1/2$.  At $\beta=1/2$ we retain the double logarithm derived in Part~I, while for $1/2<\beta<1$ the diagonal expansion remains analytic through fourth order.  None of these diagonal results is yet the complete EMT form factor.  The next section projects the known ERBL GPD onto the second moment and reconstructs the missing contribution from the reflection-odd meson spectrum.

\section{ERBL completion of the local second moment}
\label{sec:erblspectral}

The diagonal analysis determines how the wave-function boundary behavior enters the near-forward expansion, but it does not give the complete second moment.  The missing term comes from the ERBL region, where the current creates a quark--antiquark pair and the pair propagates through intermediate mesons before coupling to the external state.  The goal of this section is to project the known complete GPD onto the local second moment, identify the mesons selected by that projection, and rewrite their sum as a resolvent of the 't~Hooft Hamiltonian.  This form will make the cancellation with the diagonal contribution transparent in Sec.~\ref{sec:matching}.

The complete EMT form factor is
\begin{equation}
 \Mfull(\xi,t)=\int_{-1}^{1}\dd z\,zH_n(z,\xi,t).
 \label{eq:fullmomentdef}
\end{equation}
With the GPD and one-particle normalization adopted here, this moment is the EMT form factor,
\begin{equation}
 \Mfull(\xi,t)=\Theta_n(t),
 \label{eq:momentTheta}
\end{equation}
where the unique spin-zero EMT structure in two dimensions is~\cite{Ji:2021}
\begin{equation}
 \langle p'|T^{\mu\nu}(0)|p\rangle
 =2P^\mu P^\nu\Theta_n(t),
 \qquad
 P^\mu=\frac{p^\mu+p'^\mu}{2},
 \qquad
 \Theta_n(0)=1.
 \label{eq:EMTdecomposition}
\end{equation}
Taking the trace gives
\begin{equation}
 \langle p'|T^\mu_{\ \mu}(0)|p\rangle
 =\frac{4M_n^2-t}{2}\Theta_n(t).
 \label{eq:tracerelation}
\end{equation}
The second moment receives a diagonal contribution and an ERBL contribution,
\begin{equation}
 \Theta_n(t)=\MD(t)+\ME(t).
 \label{eq:supportdecomposition}
\end{equation}
This separation is a feature of the light-front representation, not a division of the local form factor into two observables.  Neither term is separately local, and each may contain fractional powers or logarithms.  Only their sum is the matrix element of the local EMT operator; the cancellation between the two kinematic regions must restore the analytic behavior of \(\Theta_n(t)\).

The complete light-cone GPD, including its ERBL contribution, was constructed in Refs.~\cite{Burkardt:2000,Jia:2024}.  In the ERBL region the bilocal operator couples the external meson to a two-meson intermediate sector, and the large-$N_c$ meson Hamiltonian reconnects that sector to the final state through the $1\leftrightarrow2$ transition vertex of Callan, Coote, and Gross~\cite{Callan:1976}.  The organization of form factors through confined intermediate mesons goes back to Ref.~\cite{Einhorn:1976}.  In the present normalization, the ERBL contribution is
\begin{equation}
 H_n^{\mathrm E}(z,\xi,t)=
 \theta(\xi-|z|)
 \sum_r\frac{\widehat\Gamma_{nnr}(b)}{t-\mu_r^2+i0}
 \phi_r\!\left(\frac{z+\xi}{2\xi}\right),
 \qquad
 \widehat\Gamma_{nnr}(b)
 =\sqrt{\frac{N_c}{4\pi}}\,
 \Gamma_{n;n;r}^{\rm Jia}(1-b,b),
 \label{eq:ERBLGPD}
\end{equation}
with $b=2\xi/(1+\xi)$.  The convention and sign conversion needed for the second-moment projection are collected in Appendix~\ref{app:conventions}.

Projecting Eq.~\eqref{eq:ERBLGPD} with the weight \(z\) maps the interval \(-\xi<z<\xi\) to \(0<u<1\),
\begin{equation}
 u=\frac{z+\xi}{2\xi},\qquad
 z=\xi(2u-1),\qquad \dd z=2\xi\,\dd u,
 \label{eq:ERBLchange}
\end{equation}
and introduces the local coupling
\begin{equation}
 d_r=\int_0^1\dd u\,(2u-1)\phi_r(u).
 \label{eq:dr}
\end{equation}
For an odd intermediate state the reflected-wave-function difference in the published kernel obeys
\begin{equation}
 \int_{-\xi}^{\xi}\dd z\,z\left[
 \phi_r\!\left(\frac{\xi-z}{2\xi}\right)
 -\phi_r\!\left(\frac{\xi+z}{2\xi}\right)\right]
 =-4\xi^2d_r.
 \label{eq:weightedreflection}
\end{equation}
This identity fixes the global sign of the projected source before any boundary expansion is made.
The ERBL part of the second moment is therefore
\begin{equation}
 \ME(\xi,t)=2\xi^2
 \sum_r\frac{d_r\widehat\Gamma_{nnr}(b)}{t-\mu_r^2+i0}.
 \label{eq:ERBLmoment}
\end{equation}
This equation converts the known complete GPD into the source-projected EMT moment used below.  The reflected-wave-function difference fixes the sign of the transition source, and each pole residue factorizes into the local coupling \(d_r\) and the external-state transition amplitude \(\widehat\Gamma_{nnr}\).

For equal constituent masses, Eq.~\eqref{eq:thooft} is invariant under $u\leftrightarrow1-u$.  Its eigenfunctions may therefore be chosen with definite reflection parity,
\begin{equation}
 \phi_r(1-u)=(-1)^r\phi_r(u).
 \label{eq:parity}
\end{equation}
This symmetry becomes relevant here because the local second-moment source $s(u)=2u-1$ is reflection odd.  Consequently,
\begin{equation}
 d_r=0
 \qquad
 \text{for reflection-even intermediate states},
 \label{eq:parityselection}
\end{equation}
and only the odd sector contributes to the second moment.  Isolating the intermediate wave function in the explicit transition kernel defines a source $J_n(u;b)$ by
\begin{equation}
 \widehat\Gamma_{nnr}(b)=\int_0^1\dd u\,\phi_r(u)J_n(u;b).
 \label{eq:sourcevertex}
\end{equation}
Only its odd part is relevant,
\begin{equation}
 \Jodd(u;b)=\frac12\left[J_n(u;b)-J_n(1-u;b)\right].
 \label{eq:Jodd}
\end{equation}
The resulting unprojected transition source is
\begin{align}
 J_n(u;b)={}&-2 b(1-b)
 \int_0^1\dd v\,\phi_n(v)
 \nonumber\\
 &\times
 \frac{\phi_n[b(1-u)]-\phi_n[b+(1-b)v]}
 {[bu+(1-b)v]^2}.
 \label{eq:explicitJ}
\end{align}
Equation~\eqref{eq:explicitJ} is the resulting transition source.

For the strict chiral ground state, the exact ERBL zero is already contained in the complete GPD result of Ref.~\cite{Jia:2024}.  In the present notation,
\begin{equation}
 J_0(u;b)=0,
 \qquad
 \mathcal M_{2,0}^{\mathrm E}=0.
 \label{eq:chiralzero}
\end{equation}
This exact result is stronger than the generic small-$\xi$ suppression and is kept separate from the finite-mass expansion.

The second-moment weight has now selected the odd spectrum.  Let $H_-$ denote the 't~Hooft Hamiltonian restricted to this sector.  We use $r$ for a full-spectrum label and $k=0,1,2,\ldots$ for rank within the ordered odd sector,
\begin{equation}
 H_-|-,k\rangle=\mu_{-,k}^2|-,k\rangle.
 \label{eq:oddHamiltonian}
\end{equation}
Completeness of its eigenstates converts the pole sum into the inverse-Hamiltonian matrix element
\begin{equation}
 \Gproj(t;b)=
 \sum_{r\in\mathrm{odd}}
 \frac{d_r\widehat\Gamma_{nnr}(b)}{t-\mu_r^2+i0}
 =\left\langle s\left|
 \frac{1}{t-H_-+i0}
 \right|\Jodd(b)\right\rangle.
 \label{eq:projectedresolvent}
\end{equation}
The ERBL moment is
\begin{equation}
 \ME(t)=2\xi^2(t)\,\Gproj(t;b(t)).
 \label{eq:ERBLresolvent}
\end{equation}
This representation removes the need to construct the intermediate tower state by state, while retaining its physical meromorphic structure.

The rapidity transform, parity-resolved Fredholm equation, and spectral determinants of the equal-mass Hamiltonian~\cite{Fateev:2009,Ambrosino:2025,Litvinov:2025} give the source-projected odd resolvent in the form
\begin{equation}
 \Gproj(t;b)=\frac{\mathcal N_n(t;b)}{D_-(t)},
 \label{eq:Fredholmprojected}
\end{equation}
where $D_-(t)$ is the odd-sector spectral determinant and $\mathcal N_n(t;b)$ is the corresponding augmented first minor for the sources $s$ and $J_n^-$.  A common nonzero analytic normalization of numerator and denominator leaves the quotient, residue ratios, removable-pole conditions, and off-spectrum zeros unchanged.  Appendix~\ref{app:fredholm} summarizes the source mapping and the finite-basis identities used in the numerical calculation.

Let $k=0,1,2,\ldots$ denote rank within the ordered odd sector, with eigenvalues $\mu_{-,k}^2$.  At a simple odd-sector eigenvalue,
\begin{equation}
 \res_{t=\mu_{-,k}^2}\Gproj(t;b)
 =d_{-,k}\widehat\Gamma_{nn;-,k}(b)
 =\frac{\mathcal N_n(\mu_{-,k}^2;b)}{D_-'(\mu_{-,k}^2)}.
 \label{eq:residueformula}
\end{equation}
A spectral pole is removable in the projected amplitude precisely when
\begin{equation}
 d_{-,k}\widehat\Gamma_{nn;-,k}(b)=0.
 \label{eq:removablepole}
\end{equation}
A zero of $\mathcal N_n(t;b)$ away from the spectrum is instead a destructive-interference zero among intermediate states.

The second-moment projection of the complete GPD selects the reflection-odd intermediate spectrum, converts the pole expansion generated by the transition vertex into the source-projected resolvent of $H_-$, and separates physical pole residues from removable poles and first-minor zeros.  The next section determines the near-forward projected source and its cancellation with the nonanalytic boundary terms of the diagonal overlap.

\subsection{Pairwise closure theorem}
\label{subsec:pairwise-theorem}

The diagonal calculation in Sec.~II identified the mixed coefficient $\mathcal K(p,q)$.  We can now state the corresponding closure result after the ERBL source and odd resolvent have been defined.

\paragraph*{Pairwise closure theorem.}
Let $p$ and $q$ be two distinct roots of the same equal-mass indicial equation,
\begin{equation}
 F(p)=F(q)=0,
 \qquad p\neq q,
 \qquad \sin[\pi(p+q)]\neq0,
 \qquad \sin(\pi q)\neq\sin(\pi p).
 \label{eq:pair-theorem-hypotheses}
\end{equation}
For boundary amplitudes $C_px^p$ and $C_qx^q$, the mixed diagonal contribution and the ERBL contribution generated by the same two families satisfy
\begin{equation}
 \left.\mathcal M_{2,n}^{\mathrm D}\right|_{b^{2+p+q}}
 =C_pC_q\mathcal K(p,q)b^{2+p+q},
 \qquad
 \left.\mathcal M_{2,n}^{\mathrm E}\right|_{b^{2+p+q}}
 =-C_pC_q\mathcal K(p,q)b^{2+p+q},
 \label{eq:pair-theorem-parts}
\end{equation}
so their sum vanishes at that order.  Coincident and degenerate limits are excluded from this statement and require separate limiting analyses; the canonical logarithmic case is treated separately below.

The ERBL coefficient can be derived without assuming the desired cancellation.  Let $p\ne q$ be two nonresonant roots of the same indicial equation, $F(p)=F(q)=0$, and define the odd mixed profile
\begin{equation}
 g_{pq}(u)=a_{pq}\left[u^p(1-u)^q-u^q(1-u)^p\right],
 \qquad
 a_{pq}=\frac{\sin(\pi q)-\sin(\pi p)}{\sin[\pi(p+q)]}.
 \label{eq:gpq}
\end{equation}
Expanding the exact source in Eq.~\eqref{eq:explicitJ} with the two boundary families retained gives
\begin{align}
 J_{\{p,q\}}^-(u;b)
 &=-2C_pC_qb^{p+q}\,\mathcal J_{pq}(u)+o(b^{p+q}),
 \label{eq:pairsource}\\
 \mathcal J_{pq}(u)
 &=\frac12\left[\mathcal L_{pq}(u)+\mathcal L_{qp}(u)
 -\mathcal L_{pq}(1-u)-\mathcal L_{qp}(1-u)\right],\\
 \mathcal L_{pq}(u)
 &=\FP\!\int_0^\infty\dd y\,
 \frac{y^p\left[(1-u)^q-(1+y)^q\right]}{(u+y)^2}.
 \label{eq:Lpq}
\end{align}
A direct finite-Hilbert-transform evaluation gives the operator identity
\begin{equation}
 H_-g_{pq}(u)=-2\mathcal J_{pq}(u).
 \label{eq:pairimage}
\end{equation}
The proof is given in Appendix~\ref{app:pairwise}.  The indicial conditions remove the local endpoint pieces in the action of $H_-$, while the reflection formula for the remaining beta functions fixes the coefficient $a_{pq}$.  Equations~\eqref{eq:pairsource} and \eqref{eq:pairimage} therefore imply
\begin{equation}
 H_-^{-1}J_{\{p,q\}}^-(b)
 =C_pC_qb^{p+q}g_{pq}+o(b^{p+q}).
 \label{eq:pairinverse}
\end{equation}
Projection with $s(u)=2u-1$ is now elementary:
\begin{align}
 \left\langle s\left|H_-^{-1}\right|J_{\{p,q\}}^-(b)\right\rangle
 &=2C_pC_qb^{p+q}a_{pq}
 \frac{p-q}{p+q+2}B(p+1,q+1)\notag\\
 &=2C_pC_qb^{p+q}\mathcal K(p,q).
 \label{eq:pairprojection}
\end{align}
The second equality follows from the gamma-function reflection identity and is independent of any analyticity assumption for the complete EMT form factor.  Using $2\xi^2=b^2/2+O(b^3)$ and $(t-H_-)^{-1}=-H_-^{-1}+O(b^2)$ then gives
\begin{equation}
 \left.\ME(b)\right|_{\{p,q\}}
 =-C_pC_q\,\mathcal K(p,q)\,b^{2+p+q}
 +o\!\left(b^{2+p+q}\right),
 \label{eq:pairERBL}
\end{equation}
and hence the pairwise locality identity
\begin{equation}
 \boxed{
 \left.\MD\right|_{b^{2+p+q}}
 +\left.\ME\right|_{b^{2+p+q}}=0.}
 \label{eq:paircancel}
\end{equation}
This is a dynamical consequence of the source and Hamiltonian equations, not a cancellation imposed from analyticity.

\section{Cancellation between the diagonal and ERBL contributions}
\label{sec:matching}

The preceding sections separate the origin of the nonanalytic terms from the mechanism that removes them.  The diagonal overlap supplies the boundary-sensitive powers, whereas the ERBL source propagates through the odd meson spectrum.  We now combine these two pieces along the physical elastic trajectory and follow how the complete EMT form factor recovers an ordinary near-forward expansion.  The three mass regimes differ in the form of the apparent obstruction--fractional power, logarithmic resonance, or finite fourth-order term--but the same source-projected Hamiltonian organizes all of them.

The near-forward limit is taken along the physical elastic trajectory in Eq.~\eqref{eq:kinematics}.  Thus $b\to0$ means simultaneously $\xi\to0$ and $t\to0$, and the small-$b$ expansion of the ERBL source supplies the terms needed for the slope, curvature, and cancellation of nonanalytic contributions.  To determine its power hierarchy, we first recall the leading endpoint behavior of a finite equal-mass eigenfunction,
\begin{equation}
 \phi_n(x)=C_nx^\beta+\cdots,
 \qquad
 0<\beta<1.
 \label{eq:leadingendpoint}
\end{equation}
This boundary law matters because one external wave function in the transition integral is evaluated at an argument proportional to $b$.  The limit is therefore nonuniform: a fixed-fraction expansion describes the interior, whereas the region $v=O(b)$ continues to resolve the endpoint profile.  Setting $v=by$ keeps that region finite, organizes the source by powers of $b$, and identifies the first coefficient that can survive the reflection-odd projection.  Boundary power counting first produces the candidate contribution
\begin{equation}
 \Jodd(u;b)\big|_{b^{2\beta}}
 =-C_n^2b^{2\beta}
 \left[F_\beta(u)-F_\beta(1-u)\right],
 \label{eq:leadingb2beta}
\end{equation}
where
\begin{equation}
 F_\beta(u)=\int_0^\infty\dd y\,y^\beta
 \frac{(1-u)^\beta-(1+y)^\beta}{(u+y)^2}.
 \label{eq:Fbeta}
\end{equation}
For $0<\beta<1/2$, the integral in Eq.~\eqref{eq:Fbeta} converges and can be written as a Gauss hypergeometric function.  That representation defines its analytic continuation in the parameter $\beta$ to the nonresonant range $0<\beta<1$, with the real branch for $0<u<1$ fixed by continuity from the convergence domain.  Applying the standard ${}_2F_1$ connection formula between the neighborhoods of $u=0$ and $u=1$ gives
\begin{equation}
 F_\beta(u)=F_\beta(1-u).
 \label{eq:Fbetasymmetry}
\end{equation}
The coefficient generated before antisymmetrization is therefore symmetric under $u\leftrightarrow1-u$.  Its difference in Eq.~\eqref{eq:leadingb2beta} vanishes identically, so it has no overlap with the reflection-odd second-moment channel,
\begin{equation}
 \Jodd(u;b)\big|_{b^{2\beta}}=0.
 \label{eq:b2betacancel}
\end{equation}
It cannot enter the ERBL second moment, and the first nonzero odd source must be sought at the next orders.  Appendix~\ref{app:finiteparts} displays the hypergeometric representation and the continuation formula used here.

The next candidate is the mixed boundary--bulk term, generated by the interference of the endpoint expansion of one external wave function with the finite-$v$ bulk of the other.  Its coefficient contains
\begin{equation}
 b^{1+\beta}
 \FP\!\int_0^1\dd v\,\frac{\phi_n(v)}{v^2}.
 \label{eq:mixedcandidate}
\end{equation}
To determine this coefficient, insert the endpoint expansion into the physical eigenvalue equation and examine the limit $x\to0$.  At order $x^{\beta-1}$, the explicit mass singularity cancels the corresponding singular part of the interaction; this is precisely the indicial condition in Eq.~\eqref{eq:beta} that fixes $\beta$.  Once this leading balance and the descendant fractional powers have been matched, the part of the interaction with finite $v$ contributes the $x$-independent term
\begin{equation}
 -\FP\!\int_0^1\dd v\,\frac{\phi_n(v)}{v^2}.
 \label{eq:remoteFPterm}
\end{equation}
The eigenvalue term $M_n^2\phi_n(x)$ and the remaining local terms vanish as positive powers of $x$, so consistency requires
\begin{equation}
 \FP\!\int_0^1\dd v\,\frac{\phi_n(v)}{v^2}=0.
 \label{eq:FPidentity}
\end{equation}
This ERBL-source identity is distinct from the diagonal finite-part cancellation $J_\beta=0$ derived in Part~I.  It removes the entire $b^{1+\beta}$ candidate.  Appendix~\ref{app:finiteparts} derives Eq.~\eqref{eq:FPidentity} and specifies its subtraction convention.

The two cancellations remove the apparent $b^{2\beta}$ term by reflection parity and the $b^{1+\beta}$ term by the finite-part identity.  They determine the earliest nonzero odd source in each mass regime and thus provide the bridge between the endpoint expansion and the restoration of locality by the ERBL sector.  The first surviving behavior is
\begin{equation}
 \Jodd(b)\sim
 \begin{cases}
  b^{1+2\beta},&0<\beta<\tfrac12,\\[1mm]
  b^2\ln^2(1/b),&\beta=\tfrac12,\\[1mm]
  b^2,&\tfrac12<\beta<1.
 \end{cases}
 \label{eq:Jclassification}
\end{equation}
The ERBL second moment carries the universal phase-space factor $2\xi^2=b^2/2+O(b^3)$.  Since the odd-sector resolvent is regular at $t=0$ for finite constituent mass, this factor shifts every source power upward by two powers of $b$.  The first possible ERBL contributions are therefore
\begin{equation}
 \ME(b)\sim
 \begin{cases}
  b^{3+2\beta},&0<\beta<\tfrac12,\\[1mm]
  b^4\ln^2(1/b),&\beta=\tfrac12,\\[1mm]
  b^4,&\tfrac12<\beta<1.
 \end{cases}
 \label{eq:MEclassification}
\end{equation}
This hierarchy immediately answers whether the nonvalence sector can modify the physical slope.  Because all ERBL terms start beyond $b^2$ and $t=-M_n^2b^2+O(b^3)$, for every finite equal mass
\begin{equation}
 \left.\frac{\dd\ME}{\dd t}\right|_{t=0}=0.
 \label{eq:ERBLslopenull}
\end{equation}
Thus the diagonal $b^2$ coefficient obtained in Part~I is the slope of the complete EMT form factor, while the curvature and higher derivatives still require the ERBL completion.

In the light regime $0<\beta<1/2$, the candidate $b^{2\beta}$ boundary-layer term has already been removed by reflection parity and the mixed $b^{1+\beta}$ term by the finite-part identity.  The next contribution generated by the boundary expansion is therefore $b^{1+2\beta}$; because $1+2\beta<2$, it also precedes the regular analytic $b^2$ source correction.  We factor out this power and define its coefficient function by
\begin{equation}
 \Jodd(u;b)=b^{1+2\beta}j_{\beta,n}(u)
 +o\!\left(b^{1+2\beta}\right).
 \label{eq:jbetadef}
\end{equation}
The function $j_{\beta,n}(u)$ is the nontrivial source profile that must be propagated through $H_-^{-1}$; only after this projection will its coefficient be compared with the diagonal boundary coefficient $B_{\beta,n}^{\mathrm D}$.

To evaluate that inverse-Hamiltonian projection, we seek a function $f_\beta$ whose image under $H_-$ reproduces $j_{\beta,n}$.  The operator $H_-$ preserves reflection parity, so $f_\beta$ must be odd because $j_{\beta,n}$ is odd.  Its endpoint behavior must at the same time contain the factor $[u(1-u)]^\beta$ inherited from the light boundary expansion.  These parity and endpoint constraints motivate the ansatz
\begin{equation}
 s(u)=2u-1,
 \qquad
 f_\beta(u)=[u(1-u)]^\beta s(u),
 \label{eq:fbeta}
\end{equation}
where $s(u)$ is the local second-moment weight.  Using the classical finite Hilbert transforms of Jacobi-type weights~\cite{GautschiWimp:1987}, direct evaluation then gives the source-specific Hamiltonian image relation
\begin{equation}
 \boxed{
 j_{\beta,n}=-\frac{C_n^2\rho_n}{\cos(\pi\beta)}H_-f_\beta.}
 \label{eq:Hpreimage}
\end{equation}
Equation~\eqref{eq:Hpreimage} shows that the coefficient $j_{\beta,n}(u)$ of the first surviving light-sector odd source lies in $\operatorname{Im}H_-$.  This converts the nonlocal projection $\langle s|H_-^{-1}|j_{\beta,n}\rangle$ into a local overlap with $f_\beta$.  Since $H_-$ is self-adjoint and positive for a finite equal mass~\cite{Federbush:1977,Hildebrandt:1978},
\begin{align}
 \langle s|H_-^{-1}|j_{\beta,n}\rangle
 &=-\frac{C_n^2\rho_n}{\cos(\pi\beta)}
 \int_0^1\dd u\,s^2(u)[u(1-u)]^\beta
 \notag\\
 &=-\frac{C_n^2\rho_n}{\cos(\pi\beta)}
 \frac{B(\beta+1,\beta+1)}{2\beta+3}
 =2B_{\beta,n}^{\mathrm D}.
 \label{eq:lightprojectedidentity}
\end{align}
In spectral form, inserting the complete odd-sector eigenbasis yields the exact inverse-mass-squared sum rule
\begin{equation}
 \sum_{k\geq0}
 \frac{d_{-,k}\gamma_{-,k;\beta}^{(n)}}{\mu_{-,k}^2}
 =2B_{\beta,n}^{\mathrm D},
 \qquad
 \gamma_{-,k;\beta}^{(n)}=\langle -,k|j_{\beta,n}\rangle.
 \label{eq:lightpolesum}
\end{equation}
Equation~\eqref{eq:lightprojectedidentity} fixes the inverse-Hamiltonian projection of the first surviving source coefficient, but it is not yet the ERBL moment: the latter also contains the kinematic factor $2\xi^2$ and the full resolvent $(t-H_-)^{-1}$.  Restoring these two ingredients connects the operator identity to the physical near-forward expansion.  Along the elastic trajectory,
\begin{equation}
 2\xi^2=\frac{b^2}{2}+O(b^3),
 \qquad
 (t-H_-)^{-1}=-H_-^{-1}+O(b^2).
 \label{eq:lightreconstruction}
\end{equation}
The factor $2\xi^2$ raises the $b^{1+2\beta}$ source by two powers, while the $O(t)$ correction to the resolvent contributes only at higher order.  Hence
\begin{equation}
 \ME(b)=-B_{\beta,n}^{\mathrm D}b^{3+2\beta}
 +o\!\left(b^{3+2\beta}\right).
 \label{eq:MElightcomplete}
\end{equation}
Combining this coefficient with the diagonal boundary term gives the explicit locality-restoring cancellation
\begin{equation}
 \boxed{
 \left.\MD\right|_{b^{3+2\beta}}
 +\left.\ME\right|_{b^{3+2\beta}}=0.}
 \label{eq:lightcancel}
\end{equation}
This result follows from the exact coefficient $j_{\beta,n}(u)$ of the first surviving light-sector odd source and the 't~Hooft eigenvalue equation for the external meson wave function; no analyticity assumption for the complete EMT form factor enters the derivation.  If analyticity at $t=0$ were imposed in advance, the term $b^{3+2\beta}\sim(-t)^{3/2+\beta}$ would already have to disappear from the sum of support regions.  Such an argument would establish only the necessity of cancellation.  Equations~\eqref{eq:Hpreimage}--\eqref{eq:lightcancel} determine the source profile, normalization, and inverse-mass-squared pole sum that realize it dynamically.

For the representative light mass $\mtilde^2=0.04$, the Hamiltonian image relation is confirmed both pointwise and by an independent odd-sector inversion.  A recurrence-consistent boundary matching reproduces the predicted source coefficient at the per-mille level, but is used only to check the asymptotic source.  The light curvature is extracted from the global eigenstate rather than from this boundary reconstruction.  Numerical convergence is summarized in Appendix~\ref{app:numerics}.

At $\beta=1/2$, Part~I established the first resonance between the $x^{1/2}$ and $x^{3/2}$ branches~\cite{Syamtomov:PartI}.  More generally, the multi-root boundary analysis of Ref.~\cite{Abe:1999} gives the half-integer sequence
\begin{equation}
 F\!\left(\frac12+k\right)=0,
 \qquad
 k=0,1,2,\ldots,
 \label{eq:resonanceladder}
\end{equation}
and the first generalized Frobenius term needed here is the one recalled in Eq.~\eqref{eq:canonicalendpoint}.  In the overlap region $b\ll v\ll1$, the first surviving odd source is
\begin{equation}
 \Jodd(u;b)=
 2C_nD_n(2u-1)b^2L^2
 +O(b^2L),
 \qquad
 L=\ln\frac1b.
 \label{eq:Jdoublelog}
\end{equation}
Setting $\beta=1/2$ in the function introduced in Eq.~\eqref{eq:fbeta} gives
\begin{equation}
 f_{1/2}(u)=\sqrt{u(1-u)}\,(2u-1).
\end{equation}
This specialization is the inverse-Hamiltonian image required by the canonical source.  For the present second-moment weight, a direct finite Hilbert transform gives
\begin{equation}
 H_-f_{1/2}=2\pi\,s.
 \label{eq:canonicalpreimage}
\end{equation}
Since the finite-mass odd Hamiltonian is invertible, its zero-transfer susceptibility is therefore
\begin{align}
 \langle s|H_-^{-1}|s\rangle
 &=\frac{1}{2\pi}\int_0^1\dd u\,(2u-1)^2\sqrt{u(1-u)}\notag\\
 &=\frac{1}{2\pi}\left[B\!\left(\frac32,\frac32\right)-4B\!\left(\frac52,\frac52\right)\right]
 =\frac{1}{64}.
 \label{eq:Rs}
\end{align}
Appendix~\ref{app:canonical} gives the finite-Hilbert-transform calculation leading to Eq.~\eqref{eq:canonicalpreimage}.

To propagate the source in Eq.~\eqref{eq:Jdoublelog}, we evaluate the projected resolvent along the elastic trajectory $t=t_n(b)=O(b^2)$.  Using
$(t_n(b)-H_-)^{-1}=-H_-^{-1}+O(b^2)$ together with Eq.~\eqref{eq:Rs} gives
\begin{equation}
 \Gproj(t_n(b);b)=-\frac{C_nD_n}{32}b^2L^2+O(b^2L).
 \label{eq:Gdoublelog}
\end{equation}
Equation~\eqref{eq:Gdoublelog} is the source-projected resolvent, not yet the physical ERBL moment.  Restoring the kinematic factor in Eq.~\eqref{eq:ERBLmoment} and using $2\xi^2=b^2/2+O(b^3)$ yields
\begin{equation}
 \ME(b)\supset-\frac{C_nD_n}{64}b^4L^2.
 \label{eq:MEdoublelog}
\end{equation}
The diagonal overlap derived in Part~I contains the equal and opposite term,
\begin{equation}
 \MD(b)\supset+\frac{C_nD_n}{64}b^4L^2.
 \label{eq:MDdoublelog}
\end{equation}
Consequently, the leading canonical nonanalyticity cancels in the complete EMT form factor:
\begin{equation}
 \left.\MD\right|_{b^4L^2}
 +\left.\ME\right|_{b^4L^2}=0.
 \label{eq:doublelogcancel}
\end{equation}

The next step tests the completion at single-logarithmic order and then isolates the finite $b^4$ coefficient required for the curvature.  Retaining the diagonal result through fourth order gives
\begin{equation}
 \MD(b)=1+A_{2,n}^{\mathrm D}(b^2+b^3)
 +b^4\left[d_{2,n}L^2+d_{1,n}L+d_{0,n}\right]+o(b^4).
 \label{eq:MDcanonicalfull}
\end{equation}
The combination $b^2+b^3$ is fixed by the expansion of the elastic relation $t_n(b)=-M_n^2b^2/(1-b)$; the logarithms are the remaining canonical boundary terms to be matched by the ERBL sector.

The single-logarithmic coefficient is obtained by analytically regulating the resonant endpoint pair.  Continuing the descendant as $x^{3/2}\to x^{3/2+\epsilon}$ generates $x^{3/2}\ln x$ through differentiation with respect to $\epsilon$; the finite part of the regulated beta-function expression then combines the logarithmic branch proportional to $D_n$ with the accompanying nonlogarithmic branch proportional to $K_n$.  The result is
\begin{equation}
 d_{1,n}=C_nD_n
 \left(\frac{\ln2}{16}-\frac{19}{384}\right)
 -\frac{C_nK_n}{32}.
 \label{eq:d1}
\end{equation}
This analytic regularization resolves the canonical endpoint resonance; it does not assume analyticity of the complete EMT form factor.  Appendix~\ref{app:canonical} records the regulated expansion and an independent sine-basis extraction of $K_n$.

To determine whether the diagonal $b^4L$ term is canceled, and to retain the finite term needed afterward, the ERBL moment is required through $O(b^4)$.  Since $2\xi^2=O(b^2)$ and the first resolvent correction is $O(t)=O(b^2)$, this accuracy requires only the order-$b^2$ logarithmic hierarchy of the odd source.  We therefore write
\begin{equation}
 \Jodd(u;b)=b^2\left[j_{2,n}(u)L^2+j_{1,n}(u)L+j_{0,n}(u)\right]+o(b^2),
 \label{eq:Jcanonicalexpansion}
\end{equation}
with
\begin{equation}
 j_{2,n}(u)=2C_nD_n(2u-1).
 \label{eq:j2}
\end{equation}
The ERBL moment does not require these coefficient functions pointwise; it depends only on their zero-transfer responses in the channel selected by $s(u)=2u-1$.  We collect those scalar projections as
\begin{equation}
 R_{k,n}=\langle s|H_-^{-1}|j_{k,n}\rangle.
 \label{eq:Rkn}
\end{equation}
They are, respectively, the double-logarithmic, single-logarithmic, and finite responses of the odd sector.  Substitution into the resolvent representation gives
\begin{equation}
 \ME(b)=-\frac{b^4}{2}
 \left[R_{2,n}L^2+R_{1,n}L+R_{0,n}\right]+o(b^4).
 \label{eq:MEcanonicalfull}
\end{equation}
The source-specific single logarithm can be projected without determining $j_{1,n}(u)$ pointwise.  Insert the regulated boundary pair $x^{1/2}$ and $x^{3/2+\epsilon}$ into the exact source and use Eq.~\eqref{eq:canonicalpreimage}.  Appendix~\ref{app:canonical} keeps both ordered cross terms explicitly and combines them into a symmetric regulated kernel.  The logarithmic part of the projected source is
\begin{equation}
 \langle s|H_-^{-1}|J_n^-(b)\rangle_{\rm log}
 =4C_nb^2\,\FP_{\epsilon=0}
 \left(D_n\frac{\partial}{\partial\epsilon}+K_n\right)
 \left[b^\epsilon\overline{\mathscr H}(\epsilon)\right],
 \label{eq:canonicalprojectedregulator}
\end{equation}
where $\mathscr H(p,q)$ is the ordered monomial-pair kernel defined in Appendix~\ref{app:canonical}, and
\begin{equation}
 \overline{\mathscr H}(\epsilon)=\frac12\left[\mathscr H\!\left(\frac12,\frac32+\epsilon\right)+\mathscr H\!\left(\frac32+\epsilon,\frac12\right)\right]
 =\frac{1}{64\epsilon}+h_0+O(\epsilon),
 \qquad
 h_0=\frac{19}{768}-\frac{\ln2}{32}.
 \label{eq:Hcanonicalexpansion}
\end{equation}
Since $b^\epsilon=e^{-\epsilon L}$, the finite parts are
\begin{equation}
 \FP\frac{\partial}{\partial\epsilon}
 [b^\epsilon\overline{\mathscr H}(\epsilon)]
 =\frac{L^2}{128}-h_0L+O(1),
 \qquad
 \FP[b^\epsilon\overline{\mathscr H}(\epsilon)]
 =-\frac{L}{64}+O(1).
\end{equation}
Therefore
\begin{equation}
 R_{2,n}=\frac{C_nD_n}{32},
 \qquad
 R_{1,n}=-4C_nD_nh_0-\frac{C_nK_n}{16}=2d_{1,n}.
 \label{eq:singlelogmatch}
\end{equation}
The diagonal overlap contains the single-logarithmic term $d_{1,n}b^4L$, while Eq.~\eqref{eq:MEcanonicalfull} gives the ERBL contribution $-R_{1,n}b^4L/2$.  Since Eq.~\eqref{eq:singlelogmatch} gives $R_{1,n}=2d_{1,n}$, the coefficient of $b^4L$ in the complete EMT form factor vanishes:
\begin{equation}
 \left.
 (\MD+\ME)
 \right|_{b^4L}
 =\left(d_{1,n}-\frac12R_{1,n}\right)b^4L=0.
 \label{eq:singlelogcancel}
\end{equation}
This removes the $t^2\ln(-t)$ term that would otherwise obstruct the Taylor expansion of the local form factor at $t=0$.  The cancellation follows from the projected ERBL source and is not imposed as an analyticity assumption.  If
\begin{equation}
 \gamma_{-,k;1}^{(n)}=\langle -,k|j_{1,n}\rangle,
 \label{eq:gamma1}
\end{equation}
then completeness gives the source-weighted inverse-mass-squared sum rule
\begin{equation}
 \sum_{k\geq0}
 \frac{d_{-,k}\gamma_{-,k;1}^{(n)}}{\mu_{-,k}^2}
 =C_nD_n
 \left(\frac{\ln2}{8}-\frac{19}{192}\right)
 -\frac{C_nK_n}{16}.
 \label{eq:singlelogsumrule}
\end{equation}
The left-hand side is a weighted spectral sum: each intermediate state enters through both the physical operator matrix element and the corresponding transition-source overlap.

The canonical completion formulas reduce the logarithmic coefficients, and ultimately the finite fourth-order term, to the boundary parameters $C_n$, $D_n$, and $K_n$ of the external-state wave function.  To evaluate these inputs for the low-lying states and check that they are not artifacts of a local endpoint fit, they were extracted independently by matching the sine-series representation of the numerical eigenfunctions to the canonical boundary expansion.  Table~\ref{tab:canonicalendpoint} lists the resulting coefficients together with $d_{2,n}$ and $d_{1,n}$.  The exact relation between $C_n$ and $D_n$ is imposed in the fit; varying the fit interval and basis size gives the quoted uncertainty.
\begin{table}[!htbp]
\caption{Canonical boundary coefficients and diagonal logarithmic coefficients.  The convention is $\Lambda=1$.}
\label{tab:canonicalendpoint}
\begin{ruledtabular}
\begin{tabular}{crrrrr}
$n$ & $C_n$ & $D_n$ & $K_n$ & $d_{2,n}$ & $d_{1,n}$\\
\hline
0 & $1.717046$ & $-0.843713$ & $0.652(1)$ & $-0.0226358$ & $-0.02605(1)$\\
1 & $2.648569$ & $-3.096585$ & $0.058(7)$ & $-0.128149$ & $0.0457(6)$\\
2 & $3.315511$ & $-6.074371$ & $-2.79(2)$ & $-0.314682$ & $0.413(3)$\\
3 & $3.873521$ & $-9.686534$ & $-7.32(8)$ & $-0.586266$ & $1.12(1)$\\
\end{tabular}
\end{ruledtabular}
\end{table}

The regulated projection is also consistent with a direct evaluation of the unexpanded source: over the first four states it reproduces the sign and normalization of $R_{1,n}$ within $6.5\%$.  This check is diagnostic only and is not used to determine the coefficient.  The analytic matching therefore establishes the cancellation of both canonical logarithms.  We can therefore isolate the finite fourth-order coefficient of the complete EMT form factor needed for the curvature.  Adding Eqs.~\eqref{eq:MDcanonicalfull} and \eqref{eq:MEcanonicalfull} gives
\begin{equation}
 C_{4,n}=d_{0,n}-\frac12R_{0,n}.
 \label{eq:C4}
\end{equation}
Here $C_{4,n}$ is the finite $b^4$ coefficient of the complete EMT form factor, not an additional independent parameter.

Once the nonanalytic terms have canceled, analyticity of the massive complete EMT form factor at $t=0$ gives the Taylor expansion
\begin{equation}
 \Theta_n(t)=1+\Theta_n'(0)t+\frac12\Theta_n''(0)t^2+O(t^3).
 \label{eq:Thetaseries}
\end{equation}
Matching this series to the complete $b$ expansion with
$t=-M_n^2(b^2+b^3+b^4+\cdots)$ gives
$A_{2,n}^{\mathrm D}=-M_n^2\Theta_n'(0)$ and
$C_{4,n}=A_{2,n}^{\mathrm D}+M_n^4\Theta_n''(0)/2$.  Hence
\begin{equation}
 \Theta_n''(0)=\frac{2}{M_n^4}\left[C_{4,n}-A_{2,n}^{\mathrm D}\right].
 \label{eq:curvature}
\end{equation}
The subtraction of $A_{2,n}^{\mathrm D}$ removes the $b^4$ term generated kinematically by the contribution linear in $t$.

For the light regime, the pairwise closure results remove the classified fractional powers below $b^4$.  In the massive theory, analyticity of the complete EMT form factor then implies the existence of a finite fourth-order remainder.  A formal matched representation is
\begin{equation}
 C_{4,n}^{\mathrm light}=d_{4,n}^{\mathrm D,light}
 +\mathcal I_{3,n}^{\mathrm light}\mathcal R_s(\beta),
 \label{eq:C4light}
\end{equation}
where both finite parts must be defined with the complete multi-root boundary subtraction.  The pairwise theorem guarantees cancellation of the subtraction-dependent fractional terms, but it does not by itself provide a term-by-term construction of every finite constant in the multi-root subtraction.  Equation~\eqref{eq:C4light} is therefore used here as a formal organization of the finite remainder, not as an independently evaluated analytic formula.

We instead calculate the complete EMT form factor directly from the exact diagonal overlap and the finite-basis odd-sector resolvent, using the same numerical eigenstate at each resolution.  Define
\begin{equation}
 Z_n(t)=\frac{2[\Theta_n(t)-1-\Theta_n'(0)t]}{t^2},
 \qquad Z_n(t)=\Theta_n''(0)+O(t).
 \label{eq:lightZt}
\end{equation}
The well-conditioned $(N_B,N_q)=(36,1600)$ calculation defines the central light-sector estimates.  The larger-basis but quadrature-sensitive $(40,2000)$ calculation and the lower $(32,1200)$ calculation are retained as systematic and diagnostic variations, respectively.  Two complementary extrapolations are performed: fixed-$b$ windows and state-dependent $b$ values chosen to probe common physical intervals in $|t|$.  Linear and quadratic fits are included, with the diagonal and ERBL contributions monitored separately.  The quoted central values are the medians of the accepted $(36,1600)$ fits, while the displayed ranges are enlarged to include the full $(36,1600)$--$(40,2000)$ method scan.  They are not covariance-based best-fit estimators or statistical errors.  The resulting estimates are
\begin{equation}
 \Theta_{n}^{\prime\prime}(0)\big|_{\widetilde m^2=0.04}
 =\left\{0.0434^{+0.0082}_{-0.0218},\; 0.0721^{+0.0064}_{-0.0120},\; 0.1113^{+0.0044}_{-0.0078}\right\},
 \qquad n=1,2,3.
 \label{eq:lightcurvatures}
\end{equation}
The $n=2,3$ values are stable only within the displayed method-dependent envelope; the fixed-$b$ and common-$|t|$ intercepts retain a visible offset that does not yet show asymptotic agreement.  The $n=1$ result is provisional and carries a still larger envelope.  For $n=0$, the two fitting strategies and successive resolutions do not approach a common intercept, so no ground-state light curvature is quoted.  The normalization residuals and finite-window effective slope residuals, whose amplification in $Z_n(t)$ scales as $1/t^2$ and $1/t$, respectively, are reported explicitly in Appendix~\ref{app:numerics}.

The canonical point is the boundary between two qualitatively different expansions.  For $0<\beta<1/2$, the fractional term $b^{3+2\beta}$ appears before $b^4$; at $\beta=1/2$ it becomes resonant with fourth order and generates the logarithms treated above.  We now turn to the post-resonant regime $1/2<\beta<1$, where $b^{3+2\beta}=o(b^4)$ and the curvature can be extracted from an ordinary fourth-order coefficient.  A fixed-$x$ expansion of the exact diagonal overlap remains nonuniform near the boundary, however, so the coefficient is represented by the matched Hadamard finite part
\begin{equation}
 d_{4,n}^{\mathrm D}=\FP\int_0^1\dd x\,F_{4,n}(x),
 \label{eq:d4heavy}
\end{equation}
where $F_{4,n}(x)$ is the coefficient of $b^4$ in the fixed-$x$ expansion of the exact overlap integrand.  Its only nonintegrable term is
\begin{equation}
 F_{4,n}(x)\sim\kappa_{\beta,n}x^{2\beta-3},
 \qquad
 \kappa_{\beta,n}=\frac{C_n^2}{12}\beta(\beta-1)^2(\beta-2).
 \label{eq:kappaheavy}
\end{equation}
Thus the finite part may be evaluated as
\[
 d_{4,n}^{\mathrm D}=\lim_{\epsilon\to0^+}\left[
 \int_\epsilon^1\dd x\,F_{4,n}(x)
 -\frac{\kappa_{\beta,n}}{2-2\beta}\epsilon^{2\beta-2}\right].
\]
The divergence belongs to the termwise interior representation; the matched physical coefficient is finite.

The finite fourth-order coefficient of the complete EMT form factor also requires the ERBL response.  The candidate source terms at $b^{2\beta}$ and $b^{1+\beta}$ have already vanished by reflection antisymmetry and the finite-part identity, respectively.  In the post-resonant regime the first surviving source at the order relevant for the curvature is therefore the regular $b^2$ term,
\begin{equation}
 \Jodd(u;b)=-2\mathcal I_{3,n}s(u)b^2+o(b^2),
 \qquad
 \mathcal I_{3,n}=\FP\int_0^1\dd v\,\frac{\phi_n^2(v)}{v^3}.
 \label{eq:Jheavy}
\end{equation}
Because $\phi_n(v)\sim C_nv^\beta$, this integral contains the same boundary power $v^{2\beta-3}$.  The matched subtraction convention gives
\begin{equation}
 \mathcal I_{3,n}=\lim_{\epsilon\to0^+}\left[
 \int_\epsilon^1\dd v\,\frac{\phi_n^2(v)}{v^3}
 -\frac{C_n^2}{2-2\beta}\epsilon^{2\beta-2}\right].
 \label{eq:I3finitepart}
\end{equation}
At this order the source is proportional to the single odd function $s(u)$.  Its propagation through the full intermediate-state tower is governed by the zero-transfer susceptibility
\begin{equation}
 \mathcal R_s(\beta)=\langle s|H_-^{-1}|s\rangle
 =\sum_{k\geq0}\frac{d_{-,k}^2}{\mu_{-,k}^2}.
 \label{eq:Rsheavy}
\end{equation}
Using $(t_n(b)-H_-)^{-1}=-H_-^{-1}+O(b^2)$ gives
\[
 \Gproj(t_n(b);b)=2\mathcal I_{3,n}\mathcal R_s(\beta)b^2+o(b^2),
 \qquad
 \ME(b)=\mathcal I_{3,n}\mathcal R_s(\beta)b^4+o(b^4),
\]
where the second relation also uses $2\xi^2=b^2/2+O(b^3)$.  Adding this ERBL term to the matched diagonal coefficient yields
\begin{equation}
 \boxed{
 C_{4,n}=d_{4,n}^{\mathrm D}
 +\mathcal I_{3,n}\mathcal R_s(\beta).}
 \label{eq:C4heavy}
\end{equation}
The two terms are, respectively, the diagonal finite part and the complete odd-sector ERBL response at fourth order.

The three regimes now have a common interpretation.  In the light system the Hamiltonian-image and pairwise identities remove all classified fractional powers below $b^4$; at $\beta=1/2$ the double and single logarithms cancel; for $\beta>1/2$ the diagonal finite part and regular ERBL response form the finite coefficient in Eq.~\eqref{eq:C4heavy}.  Independent source and endpoint checks are summarized in Appendix~\ref{app:numerics}.  The resulting slopes, curvatures, and trace radii are discussed next.

\section{EMT observables and finite-\texorpdfstring{$t$}{t} spectral structure}
\label{sec:observables}

\subsection{Near-forward EMT observables}
\label{subsec:nearforward}

The analytic completion separates two physical roles.  The near-forward slope and trace radius can be written directly in terms of the external wave function, whereas curvature and finite-transfer structure retain the response of the intermediate meson tower.

Equation~\eqref{eq:ERBLslopenull} promotes the diagonal coefficient derived in Part~I to the slope of the complete form factor.  The relation $A_{2,n}^{\mathrm D}=-M_n^2\Theta_n'(0)$ then gives
\begin{equation}
 \boxed{
 \Theta_n'(0)=\frac{1}{M_n^2}
 \left[\frac14+\frac12
 \int_0^1\dd x\,x(1-x)[\phi_n'(x)]^2\right].}
 \label{eq:slope}
\end{equation}
The small-transfer counting separates the roles of the two support regions.  For finite equal constituent masses, the ERBL moment begins beyond order $b^2$: as $b^{3+2\beta}$ for $0<\beta<1/2$, as $b^4L^2$ and $b^4L$ at $\beta=1/2$, and regularly at $b^4$ for $\beta>1/2$.  It therefore has no independent term linear in $t$, and Eq.~\eqref{eq:slope} is fixed entirely by the diagonal overlap.  The ERBL tower nevertheless remains essential: its intermediate-state response cancels the fractional-power or logarithmic boundary terms of the diagonal contribution, restores the local small-$t$ expansion, and enters explicitly in the curvature and higher derivatives.

The two-dimensional EMT decomposition gives the normalized scalar trace form factor~\cite{Ji:2021}
\begin{equation}
 S_n(t)=
 \frac{\langle p'|T^\mu_{\ \mu}|p\rangle}
 {\langle p|T^\mu_{\ \mu}|p\rangle}
 =\left(1-\frac{t}{4M_n^2}\right)\Theta_n(t),
 \qquad
 S_n(0)=1.
 \label{eq:normalizedtrace}
\end{equation}
To translate the analytic near-forward slope into a spatial size observable appropriate to the single spatial dimension of the model, we use the Breit-frame profile of the normalized scalar trace form factor~\cite{Choi:2021}.  At zero energy transfer $t=-\Delta^2$, and the one-dimensional transform is
\begin{equation}
 \rho_{T,n}(z)=\int_{-\infty}^{\infty}\frac{\dd\Delta}{2\pi}
 e^{-i\Delta z}S_n(-\Delta^2).
 \label{eq:tracedensity}
\end{equation}
Its normalization follows from $S_n(0)=1$, while its second spatial moment converts the form-factor slope into the longitudinal trace radius,
\begin{equation}
 \langle z^2\rangle_{T,n}=2S_n'(0)
 =2\Theta_n'(0)-\frac{1}{2M_n^2}.
 \label{eq:radiusintermediate}
\end{equation}
Substituting the slope relation in Eq.~\eqref{eq:slope} into Eq.~\eqref{eq:radiusintermediate} gives
\begin{equation}
 \boxed{
 \langle z^2\rangle_{T,n}
 =\frac{1}{M_n^2}
 \int_0^1\dd x\,x(1-x)[\phi_n'(x)]^2.}
 \label{eq:traceradius}
\end{equation}
The kinematic $1/(4M_n^2)$ term in Eq.~\eqref{eq:slope} cancels the subtraction generated by the prefactor in Eq.~\eqref{eq:normalizedtrace}, leaving the positive wave-function-gradient expression above.
This is a longitudinal radius of the normalized scalar trace form factor.  It is not a transverse gravitational radius, a pressure radius, or the radius of the ERBL pole term alone.  It also must not be confused with separate quark-mass and Coulomb radii: the DGLAP--ERBL division is kinematic, whereas the mass--interaction division is a decomposition by local operators.

Table~\ref{tab:observables} collects the complete slopes, curvatures, and longitudinal trace radii.  The canonical and heavy curvatures are stable and increase with excitation.  The light $n=2,3$ estimates follow the same trend within the current method envelope, the $n=1$ value is provisional, and the ground-state curvature is omitted because its $t^2$ coefficient is unresolved.
\begin{table}[!htbp]
\caption{Complete slopes, curvatures, and longitudinal trace radii for the light, canonical, and heavy systems in the convention $\Lambda=1$.  In general the three quantities carry units $\Lambda^{-1}$, $\Lambda^{-2}$, and $\Lambda^{-1/2}$, respectively.}
\label{tab:observables}
\begin{ruledtabular}
\begin{tabular}{lcrrr}
System & $n$ & $\Theta_n'(0)$ & $\Theta_n''(0)$ & $\sqrt{\langle z^2\rangle_{T,n}}$\\
\hline
Light &0&0.35451&---&0.36442\\
&1&0.19758&$0.043^{+0.008}_{-0.022}$&0.57433\\
&2&0.24764&$0.072^{+0.006}_{-0.012}$&0.68148\\
&3&0.31533&$0.111^{+0.004}_{-0.008}$&0.78158\\
Canonical &0&0.0811194&0.010770(5)&0.305787\\
&1&0.1527692&0.024895(5)&0.525976\\
&2&0.2339121&0.050072(15)&0.670365\\
&3&0.3157999&0.08626(7)&0.786190\\
Heavy &0&0.0282311&0.0010750415(21)&0.223110\\
&1&0.0911581&0.0060601039(12)&0.420735\\
&2&0.1652926&0.0171532660(67)&0.570979\\
&3&0.2435234&0.03619455(19)&0.694988\\
\end{tabular}
\end{ruledtabular}
\end{table}

\begin{figure}[!htbp]
 \centering
 \includegraphics[width=0.98\textwidth]{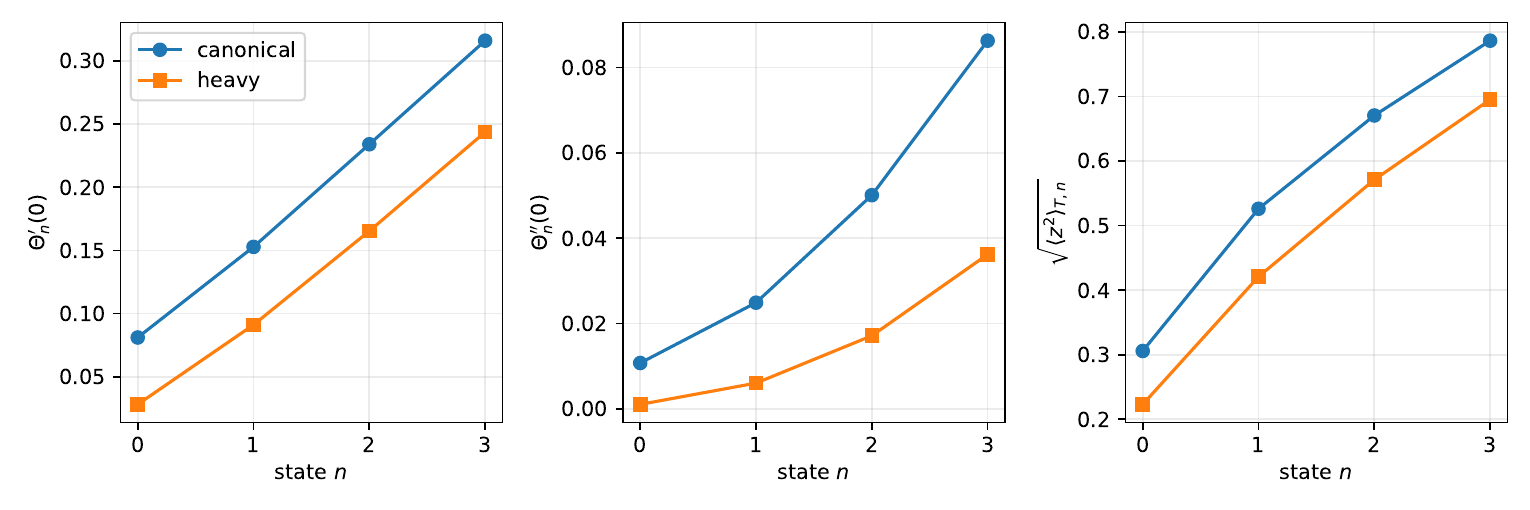}
 \caption{Complete EMT slope, curvature, and longitudinal trace radius for the canonical and heavy equal-mass systems in the convention $\Lambda=1$.  In general $\Theta_n'(0)$, $\Theta_n''(0)$, and $\sqrt{\langle z^2\rangle_{T,n}}$ carry units $\Lambda^{-1}$, $\Lambda^{-2}$, and $\Lambda^{-1/2}$, respectively.  Excitation increases all three diagnostics, whereas the heavy ground state is more compact longitudinally.}
 \label{fig:emtobservables}
\end{figure}

\subsection{Source-resolved pole structure}
\label{sec:polestructure}
\label{subsec:poles}

The numerical analysis uses the same three equal-mass benchmarks throughout: light $(\mtilde^2,\beta)=(0.04,0.10983)$, canonical $(1,1/2)$, and heavy $(16,0.93824)$.  They sample the fractional, resonant, and regular fourth-order regimes, respectively, and allow the intermediate-state response to be compared across constituent masses and external excitations.

The generic pole expansion is given in Refs.~\cite{Einhorn:1976,Burkardt:2000}.  Burkardt also argued that the lightest state with the appropriate quantum numbers should dominate at small transfer~\cite{Burkardt:2000}.  We test that expectation for the EMT second-moment source as the constituent mass and external excitation are varied.  Equivalent direct, spectral, cross-interval, and Fredholm evaluations agree at fixed discretization; the associated closure tests are summarized in Appendices~\ref{app:fredholm} and \ref{app:numerics}.

The spectral representation then asks how the ERBL response is distributed among odd intermediate mesons.  A pole is visible only if its state couples both to the local projector $s(u)=2u-1$ and to the transition source generated by the external meson.  Ordering the odd states by increasing eigenvalue, we factor the residue of the $k$th state as
\begin{equation}
 R_{-,k}^{(n)}(b)=d_{-,k}\widehat\Gamma_{nn;-,k}(b),
 \qquad
 d_{-,k}=\langle s|-,k\rangle,
 \qquad
 \widehat\Gamma_{nn;-,k}(b)=\langle -,k|\Jodd(b)\rangle.
 \label{eq:residuefactorization}
\end{equation}
The first factor is an operator overlap fixed by the intermediate state; the second contains the external-state and transfer dependence.  Their product determines whether the pole appears in the projected amplitude and with what sign and strength.  This source-resolved factorization permits tests of lowest-state dominance, broad pole saturation, removable poles, and destructive interference.

Along the physical spacelike trajectory $t=t_n(b)=-M_n^2b^2/(1-b)<0$, no spectral pole is crossed.  The partial reconstruction from the lowest $N+1$ odd states is
\begin{equation}
 S_N^{(n)}(b)=\sum_{k=0}^{N}
 \frac{R_{-,k}^{(n)}(b)}{t_n(b)-\mu_{-,k}^2},
 \label{eq:cumulative}
\end{equation}
and completeness implies $S_N^{(n)}(b)\to\Gproj(t_n(b);b)$ as the resolved tower is enlarged.  We use two complementary diagnostics,
\begin{equation}
 f_N^{\mathrm{signed}}=\frac{S_N^{(n)}}{\Gproj(t_n(b);b)},
 \qquad
 f_N^{\mathrm{abs}}=
 \frac{\sum_{k\leq N}|R_{-,k}^{(n)}(b)/(t_n(b)-\mu_{-,k}^2)|}
 {\sum_k|R_{-,k}^{(n)}(b)/(t_n(b)-\mu_{-,k}^2)|}.
 \label{eq:saturation}
\end{equation}
The signed ratio measures reconstruction of the net physical amplitude and retains all cancellations; it can be nonmonotonic, overshoot unity, or become ill-conditioned near an interference zero.  The absolute ratio instead gives the cumulative fraction of the total absolute pole weight.  It is monotonic and bounded by unity, but it is not the relative error of the signed partial sum.  In the exact definition the denominator contains the full odd tower; in the numerical tables it is evaluated with all 32 resolved odd states.  We denote by $N_{90}$ the number of lowest poles needed to reach $90\%$, so in the indexing of Eq.~\eqref{eq:cumulative} the criterion is $f_{N_{90}-1}^{\mathrm{abs}}\geq0.90$.

Table~\ref{tab:polesaturation} summarizes the scan at $b=0.25$ using 32 resolved odd states.  The light system is strongly lowest-pole dominated: the first odd state carries about $97$--$99\%$ of the absolute sum.  At the canonical point the first pole carries about $81$--$85\%$, and two poles are sufficient for $90\%$ saturation.  The heavy system is qualitatively different.  For the ground state the first pole carries only $42\%$, and nine odd states are needed for $90\%$ saturation; for $n=2$ the corresponding numbers are $35\%$ and eleven states.  The basis scan in Appendix~\ref{app:numerics} shows that $f_0^{\rm abs}$ and $N_{90}$ are stable, whereas the rank required for $99\%$ saturation continues to move outward as increasingly small high-state contributions are resolved.  Thus increasing the constituent mass does not sharpen lowest-pole dominance.  It distributes the local EMT source over a broader set of intermediate states.

\begin{table}[!htbp]
\caption{Absolute pole saturation of $\Gproj$ at $b=0.25$ on the physical spacelike trajectory.  $N_{90}$ is the number of lowest odd poles required to reach $90\%$ of the absolute spectral sum.  The entries use $(N_B,N_q)=(64,1700)$.  The light values are used only to establish the saturation pattern; canonical and heavy basis convergence is documented in Appendix~\ref{app:numerics}.}
\label{tab:polesaturation}
\begin{ruledtabular}
\begin{tabular}{lcccc}
System & $f_{0,\,n=0}^{\rm abs}$ & $N_{90}(n=0)$ & $f_{0,\,n=2}^{\rm abs}$ & $N_{90}(n=2)$\\
\hline
Light, $\mtilde^2=0.04$ & $0.995$ & $1$ & $0.970$ & $1$\\
Canonical, $\mtilde^2=1$ & $0.807$ & $2$ & $0.849$ & $2$\\
Heavy, $\mtilde^2=16$ & $0.419$ & $9$ & $0.354$ & $11$\\
\end{tabular}
\end{ruledtabular}
\end{table}

\begin{figure}[!htbp]
 \centering
 \includegraphics[width=0.485\textwidth]{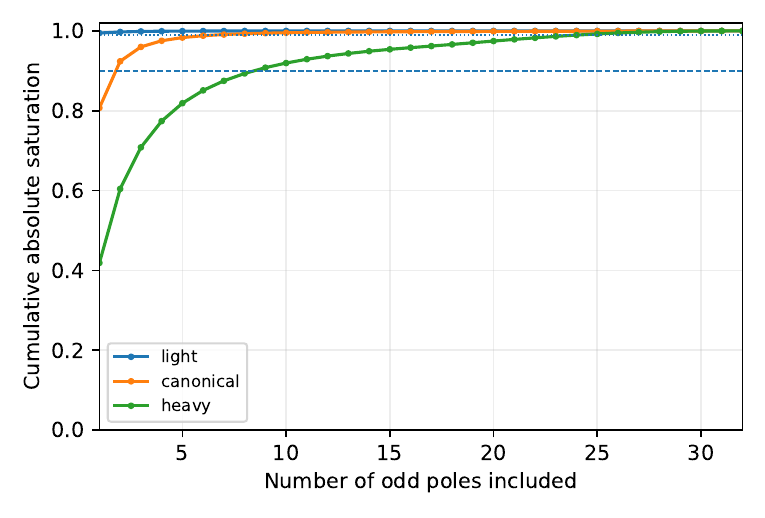}\hfill
 \includegraphics[width=0.485\textwidth]{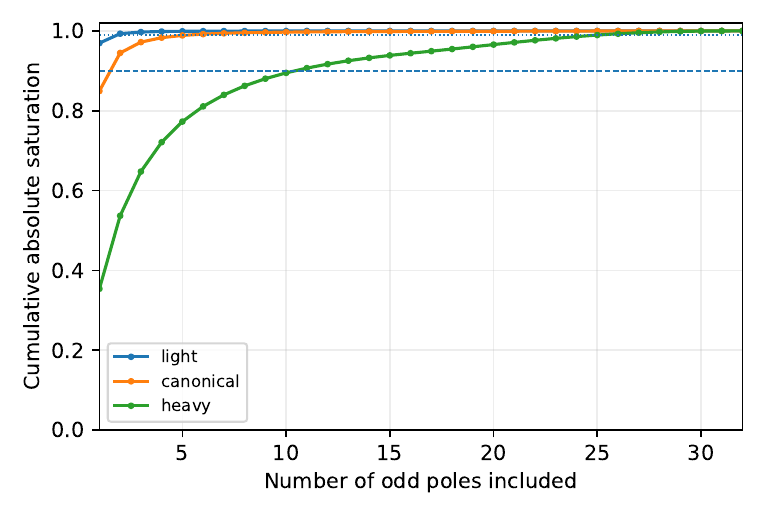}
 \caption{Cumulative absolute saturation of the odd intermediate-state tower at $b=0.25$ for the external ground state (left) and $n=2$ state (right).  The horizontal lines mark $90\%$ and $99\%$ saturation.  The $90\%$ crossing is stable in the resolved towers; the heavy $99\%$ crossing remains qualitative because its rank continues to move outward as additional small high-state contributions are resolved.  The lowest odd pole nearly exhausts the light result, two poles dominate the canonical result, while the heavy source remains distributed over much of the tower.}
 \label{fig:polesaturation}
\end{figure}

The source-projected first minor distinguishes a genuine spectral pole from a pole visible to the chosen operator and external state.  At fixed $b$, a pole at $t=\mu_{-,k}^2$ is removable when
\begin{equation}
 \widehat\Gamma_{nn;-,k}(b)=0
 \quad\Longrightarrow\quad
 R_{-,k}^{(n)}(b)=0,
 \label{eq:transitionzero}
\end{equation}
provided $d_{-,k}\neq0$.  For the external ground state, the first odd pole decouples at
\begin{equation}
 b_{\rm rem}^{(0)}=0.28873(2)
 \quad (\mtilde^2=1),
 \qquad
 b_{\rm rem}^{(0)}=0.38415706(2)
 \quad (\mtilde^2=16).
 \label{eq:groundremovable}
\end{equation}
At the roots of each discretized calculation the residuals are at machine precision.  The quoted uncertainties are instead set by the basis--quadrature changes $(N_B,N_q)=(24,300),(28,450),(32,700)$.  Excited external states possess several such zeros as $b$ varies.  They reflect sign changes of the transition source, not the mere increase of a node count.

A zero of the first-minor numerator away from the spectrum,
\begin{equation}
 \mathcal N_n(t_0;b)=0,
 \qquad D_-(t_0)\neq0,
 \label{eq:interferencezero}
\end{equation}
produces a destructive-interference zero of the projected amplitude.  Table~\ref{tab:firstminorzeros} gives the first two zeros for the ground state at $b=0.25$; their basis stability is documented in Appendix~\ref{app:numerics}.  In the canonical and heavy systems all low residues have the same sign, so one zero lies between each adjacent pair of low poles.  The zero is displaced toward the upper pole when the lower state dominates strongly.  This displacement is pronounced in the canonical system but much weaker in the heavy system, consistently with the saturation pattern above.

\begin{table}[!htbp]
\caption{First-minor zeros for the external ground state at fixed $b=0.25$.  Here $k$ is the rank within the odd sector, and the relative position is $(t_0-\mu_{-,k}^2)/(\mu_{-,k+1}^2-\mu_{-,k}^2)$.}
\label{tab:firstminorzeros}
\begin{ruledtabular}
\begin{tabular}{lcccc}
System & $\mu_{-,k}^2$ & $\mu_{-,k+1}^2$ & $t_0$ & Relative position\\
\hline
Canonical & $17.3086$ & $37.0215$ & $31.8608$ & $0.738$\\
Canonical & $37.0215$ & $56.7551$ & $53.1973$ & $0.820$\\
Heavy & $94.3765$ & $123.8249$ & $109.9049$ & $0.527$\\
Heavy & $123.8249$ & $150.1637$ & $139.8752$ & $0.609$\\
\end{tabular}
\end{ruledtabular}
\end{table}

\begin{figure}[!htbp]
 \centering
 \includegraphics[width=0.485\textwidth]{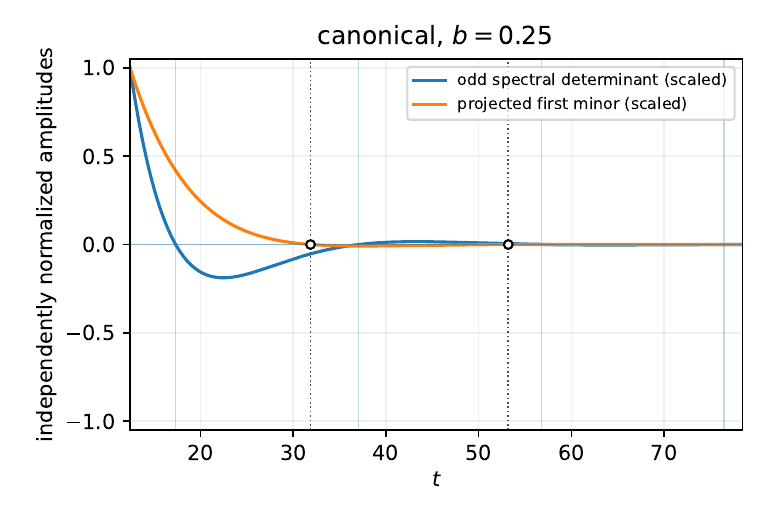}\hfill
 \includegraphics[width=0.485\textwidth]{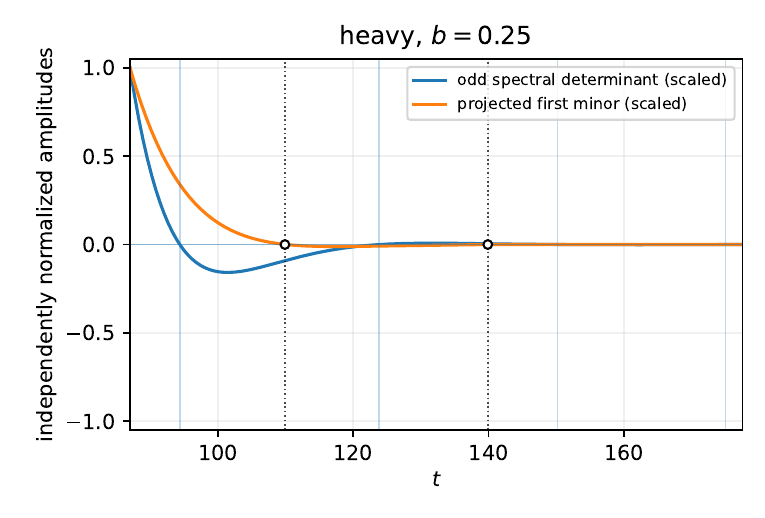}
 \caption{Independently normalized finite-basis odd spectral determinant and source-projected first minor for the canonical (left) and heavy (right) ground states at $b=0.25$.  Vertical lines mark the odd eigenvalues, while black markers and dotted lines identify the first two tabulated first-minor zeros.  A common zero of numerator and denominator would remove a pole; off-spectrum numerator zeros are destructive-interference zeros.  Independent normalization is used because only the ratio and the residue ratio are invariant.}
 \label{fig:firstminorzeros}
\end{figure}

The source dependence also produces zeros directly on the physical spacelike trajectory.  For the ground state,
\begin{equation}
 \Gproj(t_n(b);b)=0
 \quad\text{at}\quad
 (b,t)\simeq(0.2918,-0.875)
 \quad (\mtilde^2=1),
 \label{eq:physicalzerocanonical}
\end{equation}
while the heavy ground state has
\begin{equation}
 (b,t)\simeq(0.3728,-16.58).
 \label{eq:physicalzeroheavy}
\end{equation}
These zeros lie near, but not exactly at, the first-pole decoupling points in Eq.~\eqref{eq:groundremovable}, because all higher poles remain present.  Excited states show several trajectory zeros.  For example, at the highest quoted resolution the canonical $n=2$ amplitude has zeros near $b=0.081$, $0.509$, and $0.880$, while the heavy $n=2$ amplitude has zeros near $b=0.186$, $0.484$, and $0.744$.  These latter values are reported as basis estimates rather than precision observables.

\begin{figure}[!htbp]
 \centering
 \includegraphics[width=0.485\textwidth]{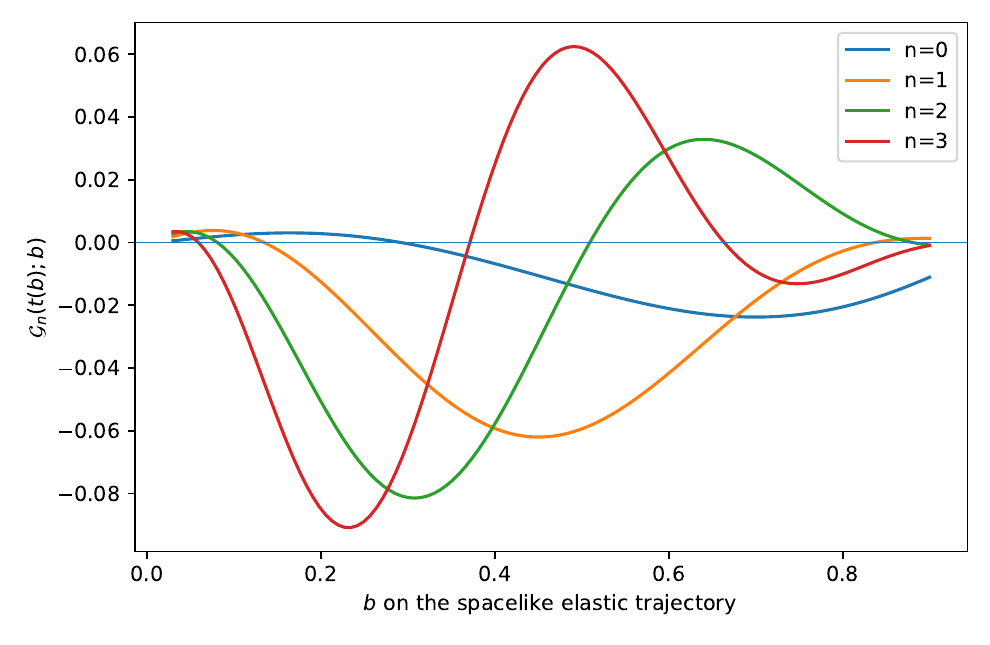}\hfill
 \includegraphics[width=0.485\textwidth]{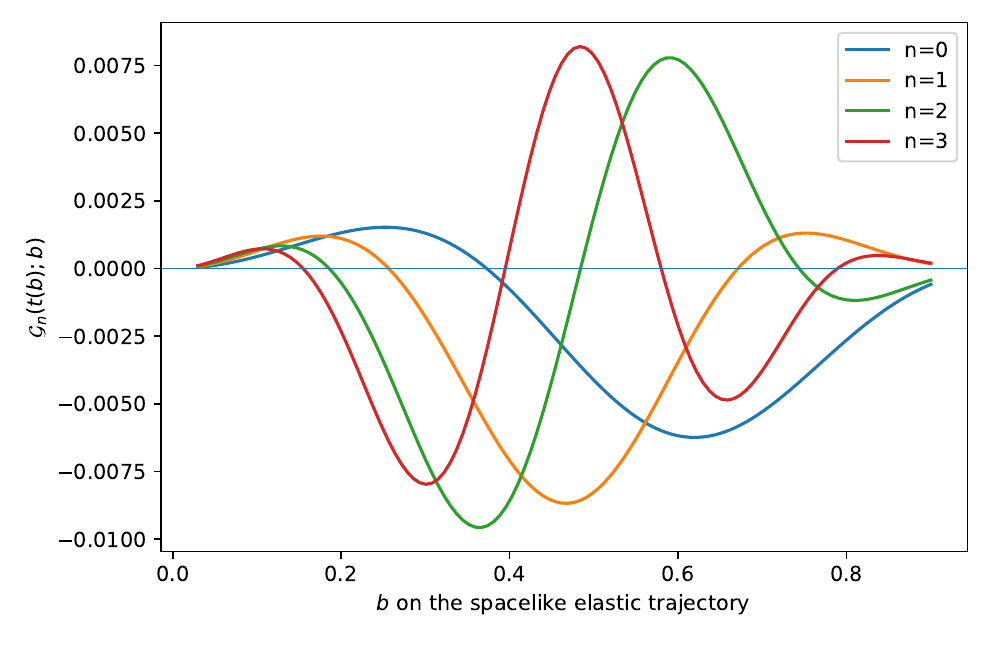}
 \caption{Projected ERBL resolvent along the physical spacelike trajectory for the first four external states in the canonical (left) and heavy (right) systems.  The zero crossings are source-interference effects of the complete tower and should not be identified with zeros of an individual transition vertex.}
 \label{fig:physicalzeros}
\end{figure}

The fixed-$b$ timelike zeros in Table~\ref{tab:firstminorzeros} are properties of a resolvent with a fixed real source.  They are not automatically zeros of the physical timelike EMT form factor.  Analytic continuation of the on-shell relation gives
\begin{equation}
 b(t)=\frac{t+\sqrt{t(t-4M_n^2)}}{2M_n^2},
 \label{eq:btanalytic}
\end{equation}
with the square-root branch chosen continuously from $t<0$.  Hence $b(t)$ is complex for $0<t<4M_n^2$ and leaves the real interval above threshold.  A physical timelike residue or zero therefore requires analytic continuation of the source itself, not substitution of a fixed real $b$ into the pole sum.  The present fixed-$b$ analysis determines pole visibility and interference unambiguously, while the spacelike trajectory analysis remains entirely on the real wave-function domain.

The pole representation applies to the complete form factor, including the direct contribution required by the local operator.  At leading large $N_c$, the timelike continuation
\begin{equation}
 t\longrightarrow t+i0
 \label{eq:continuation}
\end{equation}
produces the zero-width meson poles characteristic of the planar model~\cite{tHooft:1974,Einhorn:1976}.  Multiparticle cuts and finite widths arise beyond the approximation used here.  All direct and pole contributions must be continued on the same sheet.

\section{Conclusions}
\label{sec:conclusions}

The analysis shows that the nonanalytic behavior of the separate light-front support regions is not a pathology of the local EMT form factor.  It records how locality is distributed between the diagonal overlap and pair-creation dynamics.  In the light regime, the first surviving boundary term and the leading/next-root interference term are each generated explicitly in the diagonal overlap and removed by ERBL sources that are exact Hamiltonian images.  At the canonical point the same mechanism appears in resonant form and cancels both logarithmic orders.  For heavier constituents the cancellation is already encoded in a finite fourth-order combination.

This completion has a direct physical consequence.  The ERBL sector begins beyond linear order in $t$, so the slope of the complete EMT form factor is fixed entirely by the diagonal wave function,
\begin{equation}
 \Theta_n'(0)=\frac{1}{M_n^2}\left[\frac14+\frac12\int_0^1\dd x\,x(1-x)[\phi_n'(x)]^2\right],
\end{equation}
whereas curvature and higher derivatives require the intermediate-state tower.  The normalized scalar trace form factor therefore has the positive longitudinal radius
\begin{equation}
 \langle z^2\rangle_{T,n}=\frac{1}{M_n^2}\int_0^1\dd x\,x(1-x)[\phi_n'(x)]^2.
\end{equation}
The canonical and heavy curvatures are numerically stable.  In the light sector the second and third excited states admit method-envelope estimates stable within the current scan, the first excitation remains provisional, and the ground-state curvature is not yet resolved.

The source-projected resolvent also clarifies the finite-transfer physics.  The odd spectrum is universal, but pole visibility is controlled by the overlap of each state with both the local EMT source and the external-state transition source.  This explains why the light system is nearly lowest-pole dominated, whereas increasing the constituent mass spreads the response across a broad tower and produces removable poles and interference zeros.  The boundary analysis and spectral representation thus show how nonvalence dynamics restore locality while leaving a measurable imprint on the curvature and pole anatomy of the complete EMT form factor.

\appendix

\section{Convention matching and the complete ERBL source}
\label{app:conventions}

This appendix fixes the normalization, sign, and momentum-fraction conventions that connect the published complete GPD to the ERBL source used in the main text.  It derives the variable change from the $1\leftrightarrow2$ transition vertex to $J_n^-(u;b)$ and records the zero-transfer and chiral checks that remove otherwise ambiguous overall factors.

The complete light-cone GPD of Ref.~\cite{Jia:2024} uses meson wave functions normalized as in Eq.~\eqref{eq:thooft} and the large-$N_c$ $1\leftrightarrow2$ transition vertex of Ref.~\cite{Callan:1976}.  The vertex is evaluated at longitudinal fractions
\begin{equation}
 x_1=\frac{1-\xi}{1+\xi}=1-b,
 \qquad
 x_2=\frac{2\xi}{1+\xi}=b.
 \label{eq:appfractions}
\end{equation}
Absorbing $\sqrt{N_c/(4\pi)}$ into the definition of $\widehat\Gamma_{nnr}$ in Eq.~\eqref{eq:ERBLGPD} leaves a vertex with the same normalization as the source overlap in Eq.~\eqref{eq:sourcevertex}.

To expose the source, use the variables
\begin{equation}
 w=\frac{2b}{2-b}u,
 \qquad
 y=\frac{2(1-b)}{2-b}v.
 \label{eq:appsourcechange}
\end{equation}
The denominator and external arguments become
\begin{align}
 y+w&=\frac{2}{2-b}[bu+(1-b)v],\\
 \frac{2\xi-w}{1+\xi}&=b(1-u),\\
 \frac{2\xi+y}{1+\xi}&=b+(1-b)v,\\
 \frac{y}{1-\xi}&=v.
 \label{eq:apparguments}
\end{align}
The Jacobian cancels the common scale in $(y+w)^{-2}$ and gives the source in Eq.~\eqref{eq:explicitJ}.  The sign is fixed by the weighted intermediate-wave-function difference already displayed in Eq.~\eqref{eq:weightedreflection}:
\begin{equation}
 \int_{-\xi}^{\xi}\dd z\,z
 \left[\phi_r\!\left(\frac{\xi-z}{2\xi}\right)
 -\phi_r\!\left(\frac{\xi+z}{2\xi}\right)\right]
 =-4\xi^2d_r
 \label{eq:appzprojection}
\end{equation}
for an odd intermediate state.  Combining Eq.~\eqref{eq:appzprojection} with the kernel prefactor yields Eqs.~\eqref{eq:explicitJ} and \eqref{eq:ERBLmoment}.  No additional factor $\mathcal C_{\rm LF}$ remains.

At $b=0$ the ERBL interval has zero measure, so $\ME(0)=0$ and the diagonal normalization gives $\Theta_n(0)=1$.  For the strict chiral ground state the source itself vanishes, as stated in Eq.~\eqref{eq:chiralzero}.  These checks fix the normalization and global sign independently of the later boundary matching.

\section{Finite-part identities}
\label{app:finiteparts}

The small-$b$ source expansion contains boundary integrals whose ordinary representations diverge although their matched coefficients are finite.  This appendix specifies the common Hadamard subtraction convention, extends the diagonal identity of Part~I to the nonresonant post-canonical range, and derives the two additional identities needed specifically for the ERBL source.  These results eliminate the candidate $b^{2\beta}$ and $b^{1+\beta}$ terms before the first surviving odd source is propagated.

Part~I derived Eq.~\eqref{eq:Jbetazero} for $0<\beta<1/2$ in the diagonal overlap~\cite{Syamtomov:PartI}.  The integrand is a total derivative,
\begin{equation}
 (1+2u)u^\beta(1+u)^\beta
 =\frac{1}{\beta+1}\frac{\dd}{\dd u}
 \left[u^{\beta+1}(1+u)^{\beta+1}\right].
 \label{eq:appJbetaderivative}
\end{equation}
After the large-$u$ powers matched to the interior expansion are subtracted, the primitive has no finite constant term for $0<\beta<1$, $\beta\ne1/2$.  Its Hadamard finite part therefore vanishes, extending Eq.~\eqref{eq:Jbetazero} to the post-canonical nonresonant range used in Sec.~\ref{sec:diagonalcompletion}.  The exceptional value $\beta=1/2$ is treated by the logarithmic resonance rather than by this continuation.

For the leading ERBL source, Eq.~\eqref{eq:Fbeta} may be written
\begin{align}
 F_\beta(u)={}&(1-u)^\beta u^{\beta-1}B(\beta+1,1-\beta)\\
 &-B(\beta+1,1-2\beta)
 {}_2F_1(2,1-2\beta;2-\beta;1-u).
 \label{eq:appFhyper}
\end{align}
Equation~\eqref{eq:appFhyper} is first obtained in the convergence domain $0<\beta<1/2$.  It then defines a meromorphic continuation in $\beta$, with the branch for $0<u<1$ fixed by the original real integral.  The Gauss connection formula between the neighborhoods of $u=0$ and $u=1$, followed by the reflection and duplication identities for gamma functions, gives $F_\beta(u)=F_\beta(1-u)$ throughout the nonresonant range used in the main text.  Hence the odd $b^{2\beta}$ source vanishes before any resolvent inversion.

For $0<\beta<1$, the finite part used below is defined explicitly by
\begin{equation}
 \FP\int_0^1\dd v\,\frac{\phi_n(v)}{v^2}
 =\lim_{\epsilon\to0^+}\left[
 \int_\epsilon^1\dd v\,\frac{\phi_n(v)}{v^2}
 -\frac{C_n}{1-\beta}\epsilon^{\beta-1}\right].
 \label{eq:FPexplicit}
\end{equation}
All other root families and descendants have exponent greater than one and are integrable in this expression.  This definition also applies directly at $\beta=1/2$; no continuation from nonresonant $\beta$ is used.

Expand the full eigenvalue equation after the $x^{\beta-1}$ indicial term and all descendant fractional powers have been canceled.  The remote part of the kernel contains the analytic constant
\begin{equation}
 -\FP\int_0^1\dd v\,\frac{\phi_n(v)}{v^2}.
\end{equation}
Neither $M_n^2\phi_n(x)$ nor any remaining Frobenius family contains an independent constant.  Its coefficient must therefore vanish:
\begin{equation}
 \FP\int_0^1\dd v\,\frac{\phi_n(v)}{v^2}=0.
 \label{eq:appendixFP}
\end{equation}
This argument uses the complete family expansion and does not assume a pure-$x$ remainder.

\section{Hamiltonian image relation for the light source}
\label{app:lightpreimage}

The leading light-sector cancellation in Sec.~\ref{sec:matching} requires the inverse-Hamiltonian projection of the first nonzero odd source.  This appendix derives the source coefficient from the boundary expansion, proves that it is the image of a known odd function under $H_-$, and reduces the resulting pole sum to an elementary beta-function overlap.  It also records the operator-level and composite-source checks used to validate the identity.

After the two cancellations of Appendix~\ref{app:finiteparts}, the coefficient of $b^{1+2\beta}$ is
\begin{equation}
 j_{\beta,n}(u)=-2C_n^2\rho_nK_\beta(u),
 \label{eq:appjK}
\end{equation}
where
\begin{align}
 K_\beta(u)=\FP\int_0^\infty\dd y\,
 \frac{y^\beta}{(u+y)^2}
 \bigl\{&(1-u)^\beta(y+1-u)\\
 &-(1+y)^\beta(1+2y)\bigr\}.
 \label{eq:appKintegral}
\end{align}
A useful analytic continuation is
\begin{equation}
 K_\beta(u)=A_\beta(u)-I_0(u)-2I_1(u),
 \label{eq:appKhyper}
\end{equation}
with
\begin{align}
 A_\beta(u)={}&(1-u)^\beta\left[u^\beta B(\beta+2,-\beta)
 +(1-u)u^{\beta-1}B(\beta+1,1-\beta)\right],\\
 I_r(u)={}&B(\beta+r+1,1-2\beta-r)
 {}_2F_1(2,1-2\beta-r;2-\beta;1-u).
\end{align}

For $f_\beta(u)=[u(1-u)]^\beta(2u-1)$, integrate the subtracted kernel once by parts:
\begin{equation}
 \FP\int_0^1\dd v\,\frac{f(u)-f(v)}{(u-v)^2}
 =-\frac{f(u)}{u(1-u)}
 +\PV\int_0^1\dd v\,\frac{f'(v)}{u-v}.
\end{equation}
The boundary equation cancels the local cotangent term in the finite Hilbert transform and leaves
\begin{equation}
 H_-f_\beta=2\cos(\pi\beta)K_\beta.
 \label{eq:appendixHf}
\end{equation}
Equations~\eqref{eq:appjK} and \eqref{eq:appendixHf} prove Eq.~\eqref{eq:Hpreimage}.  Projection with $s$ reduces the inverse-Hamiltonian overlap to
\begin{equation}
 \int_0^1\dd u\,(2u-1)^2[u(1-u)]^\beta
 =\frac{B(\beta+1,\beta+1)}{2\beta+3}.
\end{equation}
In the odd spectral basis the factor $\mu_{-,k}^2$ generated by $H_-f_\beta$ cancels the inverse-mass-squared denominator state by state before completeness is used.

Equation~\eqref{eq:appendixHf} is verified independently by pointwise principal-value quadrature and by direct inversion in the odd Galerkin sector.  A recurrence-consistent composite wave function provides a separate check of the source asymptotics; the numerical comparison is summarized in Appendix~\ref{app:numerics}.

\section{Pairwise Hamiltonian identity for nonresonant indicial roots}
\label{app:pairwise}

This appendix establishes the operator identity underlying the pairwise closure theorem of Sec.~\ref{subsec:pairwise-theorem}.  Only the finite-Hilbert-transform and continuation identities needed for the reduction are retained.

To prove the mixed-family result, retain two distinct nonresonant indicial powers $C_px^p$ and $C_qx^q$, with
\begin{equation}
 F(p)=F(q)=0,\qquad p\ne q,\qquad \sin[\pi(p+q)]\ne0,\qquad \sin(\pi q)\ne\sin(\pi p).
 \label{eq:pairconditions}
\end{equation}
The last condition is a nondegeneracy requirement for the chosen Hamiltonian preimage.  Cases in which either the numerator or denominator of $a_{pq}$ vanishes require a separate limiting analysis; the canonical logarithmic pair is one important example.  The boundary scaling $v=by$ in the exact source gives Eqs.~\eqref{eq:pairsource}--\eqref{eq:Lpq}.  The ordered integral in Eq.~\eqref{eq:Lpq} has the meromorphic representation
\begin{align}
 \mathcal L_{pq}(u)
 ={}&(1-u)^q u^{p-1}B(p+1,1-p)\notag\\
 &-u^{-2}B(p+1,1-p-q)
 {}_2F_1\!\left(2,p+1;2-q;-\frac{1-u}{u}\right).
 \label{eq:Lpqhyper}
\end{align}
It is first obtained where the separate integrals converge and then continued with the real branch fixed by $0<u<1$.  This is the same Hadamard continuation used in the diagonal boundary layer.

We now derive the Hamiltonian action explicitly.  Introduce the finite Hilbert transform
\begin{equation}
 \Phi_{ab}(u)
 \equiv \PV\int_0^1\dd v\,
 \frac{v^{a-1}(1-v)^{b-1}}{u-v},
 \qquad 0<u<1.
 \label{eq:Phiab}
\end{equation}
Starting from the Euler representation outside the cut,
\begin{equation}
 \Phi_{ab}(u)=\Re\left[
 \frac{B(a,b)}{u}
 {}_2F_1\!\left(1,a;a+b;\frac1{u}+i0\right)
 \right],
 \label{eq:PhiEuler}
\end{equation}
and applying the Gauss connection formula about $z=1$, one finds
\begin{align}
 \Phi_{ab}(u)
 ={}&-\pi\cot(\pi b)u^{a-1}(1-u)^{b-1}\notag\\
 &+\frac{B(a,b-1)}{u}
 {}_2F_1\!\left(1,a;2-b;-\frac{1-u}{u}\right).
 \label{eq:PhiContinuation}
\end{align}
The first term is the real part of the cut contribution $(1-1/u\mp i0)^{b-1}$; the second term is analytic on the chosen real branch.  Equation~\eqref{eq:PhiContinuation} fixes both the branch and the sign of the local cotangent term.

For
\begin{equation}
 w_{pq}(u)=u^p(1-u)^q,
 \end{equation}
the integration-by-parts identity for the subtracted kernel gives
\begin{equation}
 H_-w_{pq}(u)
 =\frac{\widetilde m^{\,2}-1}{u(1-u)}w_{pq}(u)
 +p\Phi_{p,q+1}(u)-q\Phi_{p+1,q}(u).
 \label{eq:HwpqPhi}
\end{equation}
Substitution of Eq.~\eqref{eq:PhiContinuation} therefore yields the fully explicit expression
\begin{align}
 \mathcal A_{pq}(u)\equiv H_-w_{pq}(u)
 ={}&[\widetilde m^{\,2}-1]u^{p-1}(1-u)^{q-1}\notag\\
 &-\pi p\cot(\pi q)u^{p-1}(1-u)^q
 +\pi q\cot(\pi q)u^p(1-u)^{q-1}\notag\\
 &+\frac{pB(p,q)}{u}
 {}_2F_1\!\left(1,p;1-q;-\frac{1-u}{u}\right)\notag\\
 &-\frac{qB(p+1,q-1)}{u}
 {}_2F_1\!\left(1,p+1;2-q;-\frac{1-u}{u}\right).
 \label{eq:Apqexplicit}
\end{align}
Reflection of the interval interchanges the two exponents, so
\begin{equation}
 h_{pq}(u)=w_{pq}(u)-w_{pq}(1-u),
 \qquad
 H_-h_{pq}(u)=\mathcal A_{pq}(u)-\mathcal A_{pq}(1-u).
 \label{eq:HhpqA}
\end{equation}

It remains to compare Eq.~\eqref{eq:HhpqA} with the continued source in Eq.~\eqref{eq:Lpqhyper}.  Set $z=-(1-u)/u$ and use
\begin{align}
 \frac{\dd}{\dd z}{}_2F_1(1,p;1-q;z)
 &=\frac{p}{1-q}{}_2F_1(2,p+1;2-q;z),\label{eq:2F1derivative}\\
 B(p+1,q-1)&=\frac{p}{q-1}B(p,q),\label{eq:BetaRecurrencePair}\\
 B(p+1,1-p-q)
 &=\frac{p\sin(\pi q)}{(1-q)\sin[\pi(p+q)]}B(p,q),\label{eq:BetaReflectionPair}
\end{align}
with the corresponding relations under $p\leftrightarrow q$ and $u\leftrightarrow1-u$.  The root equations give
\begin{equation}
 \widetilde m^{\,2}-1=-\pi p\cot(\pi p)=-\pi q\cot(\pi q).
 \label{eq:rootmasspair}
\end{equation}
Applying these identities to the continued products in Eq.~\eqref{eq:Apqexplicit} yields the four source structures in Eq.~\eqref{eq:Lpqhyper}; the remaining local powers are proportional to $F(p)$ and $F(q)$ and vanish.  Direct principal-value quadrature independently verifies the resulting identity.  One obtains
\begin{align}
 &\mathcal A_{pq}(u)-\mathcal A_{pq}(1-u)\notag\\
 &\quad=-\frac{\sin[\pi(p+q)]}{\sin(\pi q)-\sin(\pi p)}
 \Bigl[
 \mathcal L_{pq}(u)+\mathcal L_{qp}(u)
 -\mathcal L_{pq}(1-u)-\mathcal L_{qp}(1-u)
 \Bigr].
 \label{eq:HhpqFullReduction}
\end{align}
Since the bracket equals $2\mathcal J_{pq}(u)$, Eqs.~\eqref{eq:HhpqA} and \eqref{eq:HhpqFullReduction} give
\begin{equation}
 H_-h_{pq}(u)
 =-2\,\frac{\sin[\pi(p+q)]}
 {\sin(\pi q)-\sin(\pi p)}\,\mathcal J_{pq}(u).
 \label{eq:Hhpqintermediate}
\end{equation}
Defining
\begin{equation}
 a_{pq}=\frac{\sin(\pi q)-\sin(\pi p)}{\sin[\pi(p+q)]},
 \qquad g_{pq}=a_{pq}h_{pq},
\end{equation}
then proves, without inference from the diagonal coefficient,
\begin{equation}
 H_-g_{pq}=-2\mathcal J_{pq}.
 \label{eq:appendixpairimage}
\end{equation}
Equations~\eqref{eq:pairsource} and \eqref{eq:appendixpairimage} give
\begin{equation}
 H_-^{-1}J_{\{p,q\}}^-(b)=C_pC_qb^{p+q}g_{pq}+o(b^{p+q}).
\end{equation}
The final projection is
\begin{align}
 \langle s|g_{pq}\rangle
 &=2a_{pq}\int_0^1\dd u\,(2u-1)u^p(1-u)^q\notag\\
 &=2a_{pq}\frac{p-q}{p+q+2}B(p+1,q+1).
 \label{eq:sgpq}
\end{align}
To compare with the diagonal coefficient, write
\begin{align}
 \mathcal K(p,q)
 &=\frac{q-p}{p+q+2}
 \left[B(p+1,-p-q-1)-B(q+1,-p-q-1)\right].
 \label{eq:Kdifference}
\end{align}
Using
\begin{equation}
 \Gamma(-p-q-1)\Gamma(p+q+2)
 =\frac{\pi}{\sin[\pi(p+q)]},
 \qquad
 \frac{1}{\Gamma(r+1)\Gamma(-r)}=-\frac{\sin(\pi r)}{\pi},
\end{equation}
one obtains
\begin{equation}
 a_{pq}\frac{p-q}{p+q+2}B(p+1,q+1)=\mathcal K(p,q).
 \label{eq:pairgammaidentity}
\end{equation}
Equations~\eqref{eq:sgpq} and \eqref{eq:pairgammaidentity} prove Eq.~\eqref{eq:pairprojection}.

The same coefficient follows independently from the diagonal boundary layer.  The ordered integral obeys
\begin{align}
 &\FP\!\int_0^\infty\dd u\,(1+2u)u^p(1+u)^q\notag\\
 &\quad=B(p+1,-p-q-1)+2B(p+2,-p-q-2)\notag\\
 &\quad=\frac{q-p}{p+q+2}B(p+1,-p-q-1)=\mathscr H(p,q).
 \label{eq:appendixpairdiagonal}
\end{align}
Adding the reversed ordering yields $\mathcal K(p,q)$.  Restoring the universal resolvent sign and $2\xi^2$ factor then proves the pairwise diagonal--ERBL cancellation at order $b^{2+p+q}$.

The proof assumes distinct nonresonant roots and the stated nondegeneracy conditions.  Coincident or degenerate limits require separate analyses; the canonical logarithms below are one such limit.  Independent Galerkin inversion and subtracted principal-value quadrature verify the operator identity without using the diagonal coefficient.

\section{Canonical logarithmic matching}
\label{app:canonical}

At $\beta=1/2$, the fractional boundary power is resonant with the regular fourth-order term and generates double and single logarithms.  This appendix supplies the regulated diagonal and ERBL projections needed to match those logarithms, derives the canonical inverse source, and documents the independent sine-basis and raw-source checks of the finite remainder.

Using the resonant boundary coefficient $D_n$ and the diagonal double-logarithmic coefficient $d_{2,n}$ from Part~I~\cite{Syamtomov:PartI}, the remaining single-logarithmic matching is controlled by the boundary regulator
\begin{equation}
 \mathscr H(p,q)=\frac{q-p}{p+q+2}B(p+1,-p-q-1)
\end{equation}
as $q=3/2+\epsilon$ gives
\begin{equation}
 \mathscr H\!\left(\frac12,\frac32+\epsilon\right)
 =\frac{1}{64\epsilon}+\frac{19}{768}-\frac{\ln2}{32}+O(\epsilon).
 \label{eq:appLaurent}
\end{equation}
Combining the finite Laurent term with the logarithmic and nonlogarithmic $x^{3/2}$ branches yields the new coefficient
\begin{equation}
 d_{1,n}=C_nD_n\left(\frac{\ln2}{16}-\frac{19}{384}\right)
 -\frac{C_nK_n}{32}.
\end{equation}

The canonical inverse source follows from one finite Hilbert transform.  With $f_{1/2}(u)=\sqrt{u(1-u)}(2u-1)$, direct evaluation gives
\begin{equation}
 \PV\int_0^1\dd v\,\frac{f_{1/2}'(v)}{u-v}
 =2\pi(2u-1).
\end{equation}
Integration by parts in the subtracted interaction gives
\begin{equation}
 \FP\int_0^1\dd v\,\frac{f_{1/2}(u)-f_{1/2}(v)}{(u-v)^2}
 =-\frac{f_{1/2}(u)}{u(1-u)}
 +\PV\int_0^1\dd v\,\frac{f_{1/2}'(v)}{u-v}.
\end{equation}
At $\widetilde m^{\,2}=1$ the local term cancels the first term on the right.  In the dimensionless normalization of Eq.~\eqref{eq:thooft}, this gives $H_-f_{1/2}=2\pi(2u-1)$ and proves Eq.~\eqref{eq:canonicalpreimage}.  The elementary integral is
\begin{equation}
 \int_0^1\dd u\,(2u-1)^2\sqrt{u(1-u)}=\frac{\pi}{32}.
\end{equation}

Now regulate the resonant descendant as $q=3/2+\epsilon$ with $p=1/2$.  Define
\begin{equation}
 \overline{\mathscr H}(\epsilon)=\frac12\left[
 \mathscr H\!\left(\frac12,\frac32+\epsilon\right)
 +\mathscr H\!\left(\frac32+\epsilon,\frac12\right)\right].
 \label{eq:symHbar}
\end{equation}
The two orderings have the same pole and constant Laurent coefficients and differ first at $O(\epsilon)$, which affects only the nonlogarithmic $b^2$ term.  The logarithmic branch is generated by $D_n\partial_\epsilon x^{3/2+\epsilon}$ and the nonlogarithmic branch by $K_nx^{3/2+\epsilon}$.  The regulated form of Eq.~\eqref{eq:pairprojection} gives
\begin{equation}
 4C_nb^2\,\FP_{\epsilon=0}
 \left(D_n\partial_\epsilon+K_n\right)
 \left[b^\epsilon\overline{\mathscr H}(\epsilon)\right].
 \label{eq:appERBLregulator}
\end{equation}
Writing $h_{-1}=1/64$ and $h_0=19/768-\ln2/32$, one finds
\begin{align}
 \FP\partial_\epsilon[b^\epsilon\overline{\mathscr H}]
 &=\frac{h_{-1}}{2}L^2-h_0L+O(1),\\
 \FP[b^\epsilon\overline{\mathscr H}]&=-h_{-1}L+O(1).
\end{align}
This proves
\begin{equation}
 R_{2,n}=\frac{C_nD_n}{32},\qquad
 R_{1,n}=-4C_nD_nh_0-\frac{C_nK_n}{16}=2d_{1,n},
\end{equation}
including the normalization and sign of the ERBL contribution.  Direct expansion of the beta and gamma functions reproduces $h_{-1}$, $h_0$, the two ordered Laurent coefficients, and $R_{1,n}=2d_{1,n}$.

The values of $K_n$ in Table~\ref{tab:canonicalendpoint} were obtained from the independent sine basis
\begin{equation}
 \phi_n(\theta)=\sum_{k=1}^{N_s}a_k^{(n)}\sin(k\theta),
 \qquad x=\frac{1-\cos\theta}{2},
\end{equation}
with $N_s=300$ and a $2500$-point quadrature.  Over $5\times10^{-4}<x<3\times10^{-2}$ the fit form was
\begin{equation}
 \frac{\phi_n(x)}{\sqrt{x}}=C_n+D_nx\ln x+K_nx
 +x^2(c_{2,n}\ln^2x+c_{1,n}\ln x+c_{0,n}),
\end{equation}
with the exact relation between $D_n$ and $C_n$ imposed.  Changes of basis size and fit window generate the quoted uncertainty.  The raw-source values in the direct-source check were obtained independently from Eq.~\eqref{eq:explicitJ} and the Galerkin inverse, without replacing the source by its regulated boundary form.  The small-$b$ scan shows the finite-resolution turnover and the absence of a controlled asymptotic plateau at the available resolution.  The complete curvatures were then extracted directly from the sum of the diagonal and ERBL contributions using Eq.~\eqref{eq:canonicalZ}.  The resolution, fit-window, and fit-order envelope is given in Appendix~\ref{app:numerics} and is used for every canonical state.

\section{Fredholm source mapping and validation}
\label{app:fredholm}

The finite-$t$ analysis in Sec.~\ref{sec:polestructure} uses a Fredholm determinant and a source-dependent first minor as an independent representation of the projected resolvent.  This appendix identifies the transformed second-moment projector and ERBL source, then reduces the construction to finite-basis determinant formulas that validate the direct solve, spectral sum, residues, and complex continuation.

We use the established rapidity transform, inhomogeneous equation, parity decomposition, and spectral determinants of Refs.~\cite{Fateev:2009,Ambrosino:2025,Litvinov:2025}, together with the standard first-minor/resolvent identity~\cite{Feinberg:2004}.  Let $\mathscr F$ denote their endpoint-weighted transform and let $A_t$ be the diagonal factor in the standard odd-sector Fredholm factorization.  For the present second-moment source, the only additional mapping is
\begin{equation}
 \widetilde s_t=A_t^{-1\dagger}\mathscr F s,
 \qquad
 \widetilde J_{n,t}=A_t^{-1}\mathscr F J_n^-.
\end{equation}
These source factors define the augmented first minor $\mathcal N_n(t;b)$ in Eq.~\eqref{eq:Fredholmprojected}.  A change of determinant regularization multiplies $D_-$ and $\mathcal N_n$ by the same nonzero analytic function, so the physical quotient and residue ratios are unchanged.

For a finite nonorthogonal basis the same statements reduce to
\begin{align}
 \Gproj&=g^T(tA_--H_-)^{-1}j,\\
 \Delta_N&=\det(tA_--H_-),\\
 \mathcal N_{n,N}&=-\det\begin{pmatrix}tA_--H_-&j\\g^T&0\end{pmatrix}.
\end{align}
The matrix determinant lemma gives the independent rank-one identity
\begin{equation}
 1+\eta\Gproj=
 \frac{\det[tA_--H_-+\eta jg^T]}{\det[tA_--H_-]}.
\end{equation}
These formulas, the spectral sum, the direct solve, and the transformed overlap were tested independently; their maximum errors are listed in Appendix~\ref{app:numerics}.  The complex-sheet row compares the direct and spectral resolvents both away from the real axis and at $t=\mu_0^2+i10^{-3}\mu_0^2$, so the same $+i0$ continuation is used in both representations.

\section{Numerical convergence}
\label{app:numerics}

This appendix summarizes the convergence tests needed to distinguish exact fixed-discretization identities from quantities requiring continuum control.  The composite light construction is used only to check source asymptotics; light curvatures are obtained from a separate global calculation and assigned state-dependent status.

The endpoint-adapted basis is
\begin{equation}
 f_k(x)=[x(1-x)]^\beta P_k^{(2\beta,2\beta)}(2x-1).
\end{equation}
The singular interaction is evaluated through the symmetric quadratic form
\begin{equation}
 \frac12\int_0^1\dd x\dd y\,
 \frac{[f_k(x)-f_k(y)][f_l(x)-f_l(y)]}{(x-y)^2}.
\end{equation}
Product quadrature includes the continuous diagonal limit
$\frac12\sum_iw_i^2f_k'(x_i)f_l'(x_i)$; omitting it produces an $O(N_q^{-1})$ spectral bias.  The independent sine basis used in Appendix~\ref{app:canonical} does not factor out the boundary exponent.

The endpoint-adapted Jacobi and independent sine solutions follow the matched boundary behavior in the canonical and heavy systems.  In the light system finite bases resolve the leading $x^\beta$ power but represent the higher indicial families nonuniformly near the boundary; the composite reconstruction is therefore used only to test the source asymptotics, not as a pointwise eigenfunction.

The light Hamiltonian-image projection converges monotonically through $N_q=2400$.  Composite reconstructions at two resolutions and three matching points satisfy the recurrence, finite-part, matching, and normalization constraints; their fitted coefficient agrees with the analytic prediction within $2.4\times10^{-3}$.  The higher-root amplitudes and pointwise residual remain matching dependent, so this construction is used only as a source-asymptotic check.  Heavy fourth-order coefficients and removable-pole locations are stable under the quoted resolution scans.  Fixed-$b$ closures test algebraic equivalence at a common discretization and are not, by themselves, continuum tests.

\begin{figure}[!htbp]
 \centering
 \includegraphics[width=0.94\textwidth]{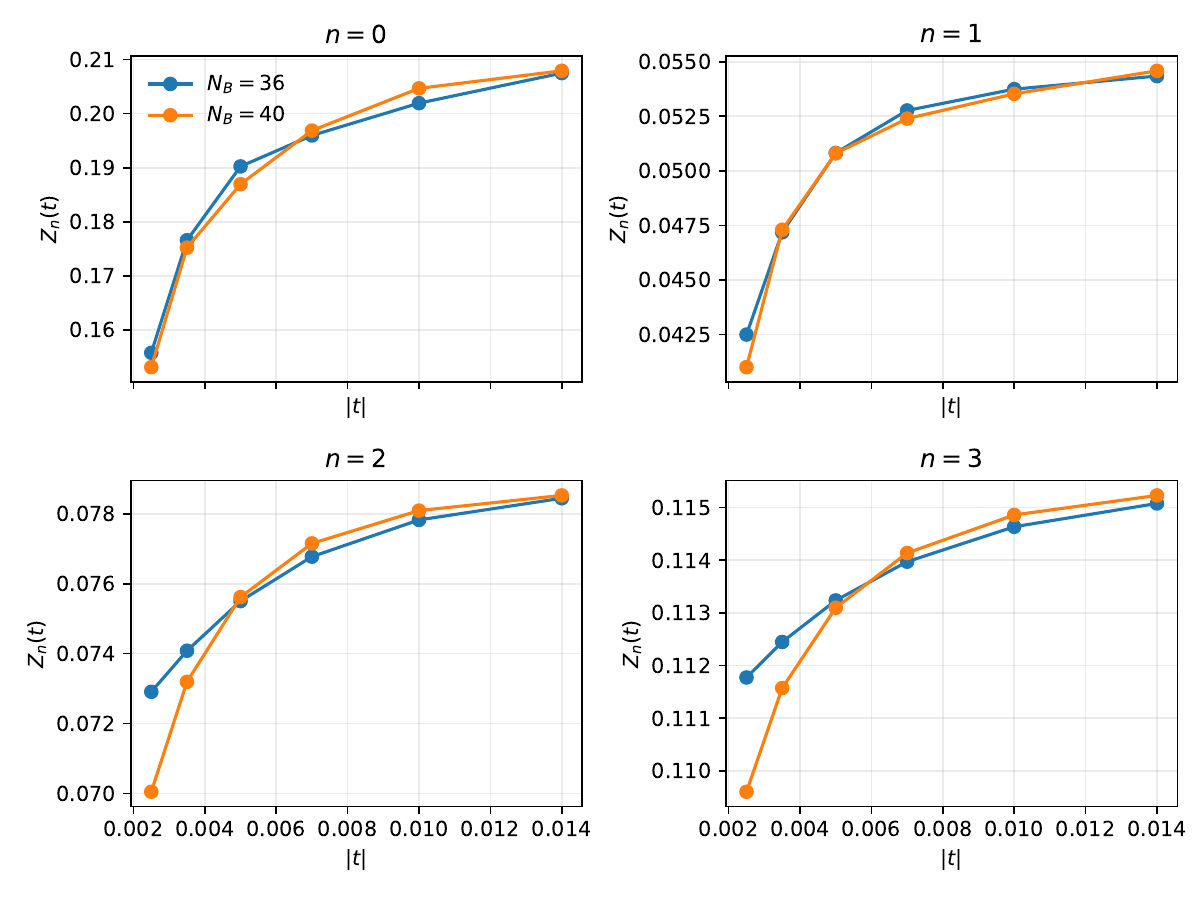}
 \caption{Direct light-sector curvature diagnostic $Z_n(t)$ at the $(36,1600)$ and $(40,2000)$ endpoint-adapted resolutions, evaluated at common physical values of $|t|$.  All quantities in each curve are computed from the same numerical eigenstate.  The $n=2,3$ curves remain inside the displayed method envelope, without demonstrated asymptotic agreement of the two window prescriptions; the $n=1$ intercept is provisional and the $n=0$ intercept is unresolved.}
 \label{fig:lightcurvatureconvergence}
\end{figure}

The mixed leading/next-root contribution $b^{2+\beta+\gamma_1}$ is isolated analytically and canceled by its ERBL counterpart, so no classified fractional boundary power remains below $b^4$.  The curvature is then extracted from the unpatched complete EMT form factor.  The central scan uses the well-conditioned $(N_B,N_q)=(36,1600)$ calculation and compares fixed-$b$ fits with state-dependent $b$ values corresponding to common physical $|t|$ intervals.  The larger-basis $(40,2000)$ calculation is retained only to broaden the systematic envelope because its normalization is more quadrature sensitive; $(32,1200)$ is a lower-resolution diagnostic.  Figure~\ref{fig:lightcurvatureconvergence} compares the $(36,1600)$ and $(40,2000)$ scan points at common physical values of $|t|$.  The $n=2,3$ curves remain within the displayed method-dependent envelope, but the fixed-$b$ and fixed-$|t|$ limits retain a visible offset.  The $n=1$ intercept has a larger scheme dependence, and the ground-state curve remains unresolved.  The smoothness of the finite-$|t|$ ground-state curve does not imply a stable intercept: subtraction of the analytic linear term and extrapolation to $t=0$ amplify small normalization, effective-slope, and fit-window differences.
\begin{table}[!htbp]
\caption{Light-curvature estimates from the direct complete EMT form factor.  Central values and scheme medians use the well-conditioned $(N_B,N_q)=(36,1600)$ calculation.  The full envelope also includes the quadrature-sensitive $(40,2000)$ variation, fixed-$b$ and common-$|t|$ windows (including smaller common-$|t|$ intervals), and linear and quadratic extrapolations.  It is a systematic scan envelope, not a statistical error.  The $n=1$ value is provisional because the two window prescriptions differ appreciably.}
\label{tab:lightcurvatureconvergence}
\begin{ruledtabular}
\begin{tabular}{ccccc}
$n$ & full envelope & $(36,1600)$ fixed-$b$ median & $(36,1600)$ fixed-$|t|$ median & status\\ \hline
0 & $-0.634$ to $0.174$ & $-0.019$ & $0.148$ & unresolved\\
1 & $0.0216$ to $0.0516$ & $0.0499$ & $0.0395$ & provisional\\
2 & $0.0601$ to $0.0785$ & $0.0778$ & $0.0708$ & stable within method envelope\\
3 & $0.1035$ to $0.1157$ & $0.1153$ & $0.1106$ & stable within method envelope\\
\end{tabular}
\end{ruledtabular}
\end{table}

At $(36,1600)$, $|\Theta_n(0)-1|\leq1.2\times10^{-12}$; at $(40,2000)$ the corresponding residuals range from $6.8\times10^{-9}$ to $5.6\times10^{-8}$, reflecting the separate quadrature sensitivity of the larger basis.  The $(40,2000)$ calculation is not treated as a higher-accuracy resolution point; it is retained only as a conservative larger-basis, quadrature-sensitive variation in the full envelope.  Quadratic fits to the common-$|t|$ points give finite-window effective slope residuals of $(9.3,0.55,1.34,0.97)\times10^{-5}$ at $(36,1600)$ and $(6.0,1.37,0.58,0.62)\times10^{-5}$ at $(40,2000)$ for $n=0,1,2,3$.  Because these finite-window residuals enter $Z_n(t)$ with $1/t$ or $1/t^2$ enhancement, they diagnose the much poorer ground-state conditioning, although they also contain higher-order fit-window effects.  They are contained in the pointwise values and hence in the observed scan spread, but are not propagated as a separate statistical uncertainty.  The diagonal and ERBL pieces were also monitored separately to ensure that their cancellation is numerically controlled.

A curvature-specific tower test was also performed at $(N_B,N_q)=(36,1600)$.  Over $|t|=0.0035$--$0.010$, retaining 8 of 18 odd states changes $Z_n(t)$ by at most $5.1\times10^{-6}$ for $n=1,2,3$, and 12 states reduce this to $2.4\times10^{-6}$.  The high-state tail is therefore well below the method-envelope uncertainty.

The pole-saturation and first-minor scans were repeated with 16--32 resolved odd states.  Table~\ref{tab:saturationconvergence} shows the endpoint resolutions.  The lowest-pole fractions are stable to better than $10^{-5}$ and $N_{90}$ is stable once at least 20 odd states are present.  The heavy $N_{99}$ rank continues to increase because the last percent is spread over many individually small high states, while the total absolute sum is already stable.
\begin{table}[!htbp]
\caption{Basis convergence of the absolute pole saturation at $b=0.25$.  Entries are shown for the smallest and largest odd towers in the scan.}
\label{tab:saturationconvergence}
\begin{ruledtabular}
\begin{tabular}{lccccc}
System, $n$ & $N_{\rm odd}$ & $f_0^{\rm abs}$ & $N_{90}$ & $N_{99}$ & $\sum_k|G_k|$\\
\hline
Canonical, 0 & 16 & 0.807205 & 2 & 7 & 0.001669315\\
             & 32 & 0.807200 & 2 & 7 & 0.001669363\\
Canonical, 2 & 16 & 0.849493 & 2 & 6 & 0.071666285\\
             & 32 & 0.849494 & 2 & 6 & 0.071665710\\
Heavy, 0     & 16 & 0.418804 & 9 & 14 & 0.0015250223\\
             & 32 & 0.418804 & 9 & 25 & 0.0015250223\\
Heavy, 2     & 16 & 0.354065 & 10 & 14 & 0.0034136367\\
             & 32 & 0.354065 & 11 & 26 & 0.0034136366\\
\end{tabular}
\end{ruledtabular}
\end{table}
The first two ground-state first-minor zeros vary by less than $1.6\times10^{-4}$ in the canonical system and $10^{-6}$ in the heavy system over the same scan.  The convergence scan confirms the quoted stability.

For the canonical system the complete EMT form factor was evaluated before fitting, so the logarithms cancel between support regions numerically.  The analytic slope contribution is subtracted in the physical variable $t$:
\begin{equation}
 Z_n(t)=\frac{2\,[\Theta_n(t)-1-\Theta_n'(0)t]}{t^2},
 \qquad Z_n(t)=\Theta_n''(0)+O(t).
 \label{eq:canonicalZ}
\end{equation}
Using the exact relation $t=-M_n^2b^2/(1-b)$ avoids truncating the kinematic series in $b$.  Linear and quadratic fits of $Z_n(t)$ were repeated at the two highest resolutions and over three $b$ windows.  Table~\ref{tab:canonicalconvergence} lists the linear/quadratic intercepts.
\begin{table}[!htbp]
\caption{Canonical curvature convergence from direct extrapolation in $t$.  Each entry gives the linear/quadratic intercept of $Z_n(t)$ over the stated $b$ window.}
\label{tab:canonicalconvergence}
\begin{ruledtabular}
\begin{tabular}{cccccc}
$(N_B,N_q)$ & $b$ window & $n=0$ & $n=1$ & $n=2$ & $n=3$\\ \hline
$(28,900)$ & $0.04$--$0.10$ & 0.0107677/0.0107654 & 0.0248930/0.0248906 & 0.0500720/0.0500747 & 0.0862600/0.0862776\\
            & $0.05$--$0.12$ & 0.0107692/0.0107653 & 0.0248939/0.0248921 & 0.0500674/0.0500747 & 0.0862372/0.0862788\\
            & $0.06$--$0.14$ & 0.0107715/0.0107687 & 0.0248941/0.0248959 & 0.0500574/0.0500780 & 0.0861903/0.0862811\\
$(32,1200)$& $0.04$--$0.10$ & 0.0107702/0.0107664 & 0.0248957/0.0248931 & 0.0500744/0.0500758 & 0.0862626/0.0862803\\
            & $0.05$--$0.12$ & 0.0107724/0.0107708 & 0.0248968/0.0248975 & 0.0500705/0.0500801 & 0.0862401/0.0862840\\
            & $0.06$--$0.14$ & 0.0107739/0.0107733 & 0.0248960/0.0248993 & 0.0500596/0.0500822 & 0.0861921/0.0862841\\
\end{tabular}
\end{ruledtabular}
\end{table}
\FloatBarrier
The canonical values quoted in Table~\ref{tab:observables} use the mean and full envelope of these twelve determinations, incorporating resolution, fit-window, and fit-order variation.  Independent direct, spectral, cross-interval, Fredholm, and bordered-minor constructions of the projected resolvent agree to at least $10^{-11}$ relative accuracy, and usually to machine precision.

\renewcommand{\bibsection}{}
\par\bigskip
\noindent\rule{\textwidth}{0.4pt}
\par\medskip
\bibliography{thooft_lightfront_diagnostics_part2}

\end{document}